\documentclass[journal]{IEEEtran}

\usepackage{pifont}
\newcommand{\cmark}{\ding{51}}
\newcommand{\xmark}{\ding{55}}

\usepackage{hyperref}
\usepackage[table,xcdraw]{xcolor}

\hypersetup{
	colorlinks=true,
  linkcolor=blue,
  urlcolor=blue,
	breaklinks=true
}

\usepackage{subfigure}
\usepackage{enumitem}

\usepackage[flushleft]{threeparttable}

\usepackage{graphicx}
\usepackage{cite}
\usepackage{booktabs}
\usepackage{orcidlink}

\begin{document}

\bstctlcite{IEEEexample:BSTcontrol}

\title{CLAD: Robust Audio Deepfake Detection Against Manipulation Attacks with Contrastive Learning}
\author{Haolin Wu$^{\orcidlink{0009-0007-0493-5013}}$, Jing Chen$^{\orcidlink{0000-0002-7212-5297}}$, Ruiying Du$^{\orcidlink{0000-0002-3634-3385}}$, Cong Wu$^{\orcidlink{0000-0002-0930-0283}}$, Kun He$^{\orcidlink{0000-0003-3472-419X}}$, \\
Xingcan Shang$^{\orcidlink{0009-0002-7640-2922}}$, Hao Ren$^{\orcidlink{0000-0001-7909-3753}}$, Guowen Xu$^{\orcidlink{0000-0002-9764-9345}}$
}

\maketitle

\begin{abstract}

  The increasing prevalence of audio deepfakes poses significant security threats, necessitating robust detection methods. While existing detection systems exhibit promise, their robustness against malicious audio manipulations remains underexplored. To bridge the gap, we undertake the first comprehensive study of the susceptibility of the most widely adopted audio deepfake detectors to manipulation attacks. Surprisingly, even manipulations like volume control can significantly bypass detection without affecting human perception. To address this, we propose CLAD~(Contrastive Learning-based Audio deepfake Detector) to enhance the robustness against manipulation attacks. The key idea is to incorporate contrastive learning to minimize the variations introduced by manipulations, therefore enhancing detection robustness. Additionally, we incorporate a length loss, aiming to improve the detection accuracy by clustering real audios more closely in the feature space. We comprehensively evaluated the most widely adopted audio deepfake detection models and our proposed CLAD against various manipulation attacks. The detection models exhibited vulnerabilities, with FAR rising to 36.69\%, 31.23\%, and 51.28\% under volume control, fading, and noise injection, respectively. CLAD enhanced robustness, reducing the FAR to 0.81\% under noise injection and consistently maintaining an FAR below 1.63\% across all tests. Our source code and documentation are available in the artifact repository (\url{https://github.com/CLAD23/CLAD}).

\end{abstract}


\section{Introduction}

\IEEEPARstart{I}{n} recent years, deep learning has made remarkable progress in speech synthesis ~\cite{wang_tacotron_2017,  huang_prodiff_2022, li_styletts_2023} and voice conversion ~\cite{li_styletts-vc_2023, tang_avqvc_2022, li_freevc_2023}. These technologies advanced the creation of highly realistic and natural-sounding speech, which enhances user interaction in various applications. However, they also pose serious risks. Malicious attackers may use these technologies to generate high-quality audio deepfakes and then exploit them for phone scams~\cite{audio_deepfake_scams_euronews}, bypassing speaker recognition~\cite{wenger_hello_2021} and spreading disinformation on the web~\cite{khanjani_audio_2022}. 
In response, researchers have developed various methods to detect audio deepfakes, including acoustic features such as spectral features~\cite{albadawy_detecting_2019}, linear frequency cepstral coefficients (LFCC) ~\cite{lavrentyeva_stc_2019} , and constant Q cepstral coefficients (CQCC) ~\cite{todisco_constant_2017}, as well as deep learning approaches such as RawNet2~\cite{tak_end--end_2021}, Res-TSSDNet\cite{hua_towards_2021} and AASIST~\cite{jung_aasist_2022}.
These methods have shown impressive performance on large-scale public datasets like the ASVspoof dataset~\cite{wang_asvspoof_2020, yamagishi_asvspoof_2021}.

\begin{figure}[!t]
  \centering
  \includegraphics[width=0.85\columnwidth]{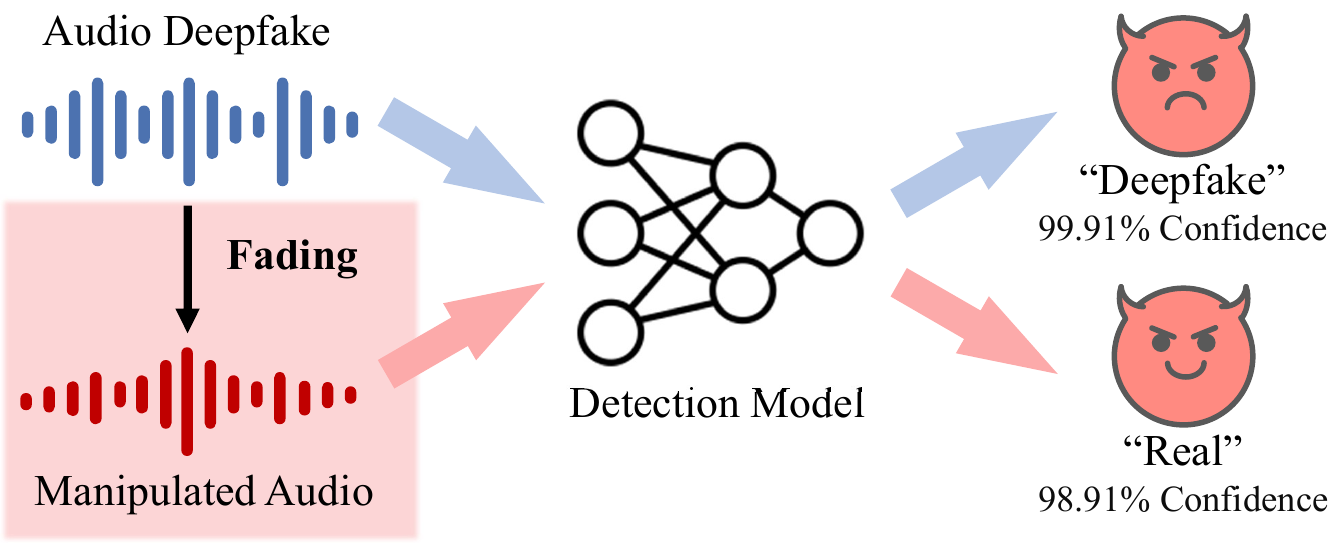}
  \caption{Illustration of the manipulation attacks. Though detection model performs well on the original audio, we found that simple manipulations like fading could bypass it. }
  \label{fig:introduction_manipulation_attacks}
\end{figure}

Despite the impressive performance in detecting audio deepfakes,  these methods' effectiveness and robustness in real-world scenarios have yet to be investigated, especially when faced with simple but natural audio manipulations.
Most existing methods are evaluated using raw output of synthesis techniques, neglecting the possible manipulations applied to the audio by the adversary~\cite{lavrentyeva_stc_2019, tak_end--end_2021, hua_towards_2021,jung_aasist_2022}.  
While the latest ASVspoof 2021 dataset~\cite{yamagishi_asvspoof_2021} introduces more realistic data subjected to telephony channel transmission and compression,
these modifications are significantly limited, and fail to reflect the techniques a real attacker might employ to fool the detection system.
Therefore, a thorough understanding of these methods' robustness against simple manipulations is essential to ensure their efficacy in mitigating deepfake threats.

To bridge the gap, we perform the first systematic large-scale study on audio manipulation attacks in the audio deepfake detection task,
where an attacker aims to deceive the detection system by simply manipulating the audio data in an imperceptible way.
For example, the attacker may apply fading to the audio to bypass the detector, as shown in Fig.~\ref{fig:introduction_manipulation_attacks}.
There have been previous works ~\cite{zhang_black-box_2020, liu_adversarial_2019,panariello_malafide_2023} that use adversarial attacks to evade detection systems, these attacks are costly and often require prior knowledge of the target system. In contrast, the manipulation attack is more realistic and cheaper, posing a serious threat to detection systems. Specifically, we design 7 manipulations, i.e., noise injection, volume control, fading, time stretching, resampling, time shifting, echoes adding, and investigate the robustness of widely adopted detection methods against these manipualtions. We observe that these methods fail to counteract the manipulation attacks and the performance drops significantly under these attacks.

In this paper, we propose a Contrastive Learning-based Audio deepfake Detector (CLAD) to enhance the robustness against audio manipulation attacks.
The key idea is to use contrastive learning to train a robust audio encoder that effectively differentiates audio deepfakes by learning intrinsic features.
By training the encoder to produce similar feature representations for the same audio under different augmentations, and dissimilar representations for different audios, the encoder can learn robust features that are more resistant to manipulation attacks.
To further enhance the effectiveness of CLAD, we developed length loss, which compels the encoder to produce short feature vectors for real audios and long feature vectors for audio deepfakes.
By clustering the features of real audios around the origin of the high-dimensional space and pushing the features of deepfakes away, the length loss effectively utilizes label information to improve the detection performance of CLAD.
The key insight is that real audios exhibit more resemblance to one another than audio deepfakes, which are generated from distinct synthesis methods.

In summary, our paper makes the following contributions:
\begin{itemize}[leftmargin=*]
  \item We perform the first comprehensive study on the robustness of widely adopted audio deepfake detection methods against audio manipulation attacks and expose the significant threat of manipulation attacks.
        We find that these detection methods are vulnerable to simple manipulation attacks, such as volume control and fading.
  \item To the best of our knowledge, we present the first audio deepfake detection method that is robust to manipulation attacks, namely CLAD. To achieve this, we introduce contrastive learning, which learns a robust encoder to produce reliable features for different audio manipulations.
        To enhance the accuracy, we also design length loss as a new learning strategy to enhance the encoder.
  \item We conduct extensive experiments to evaluate the effectiveness of CLAD in defending against manipulation attacks under different scenarios.
        The results suggest that CLAD is effective and robust against manipulation attacks. e.g.,  achieving an overall FAR of less than 1.63\%
        across all manipulations. CLAD also shows the significant improvements over existing methods.

\end{itemize}

\section{Manipulation Attacks}
\label{sec:manipulation_attacks}

\subsection{Manipulations}

To attack the audio deepfake detection systems, adversaries attempt to manipulate the audio with simple and common methods.
We discuss many kinds of manipulation approaches as never before, some of which have received little attention in prior work~\cite{wang_deepsonar_2020, hojjati_self-supervised_2022}.

\textbf{Noise injection} applies additive noise to the audio signal, which is a common occurrence in real-world environments.
We consider two typical noise types, gaussian white noise (WN) and environmental noise (EN).
For our experimental evaluation, we consider the following environmental noises: wind, footsteps, breathing, coughing, rain, clock tick and sneezing. We control the strength of the noise by adjusting the signal-to-noise ratio (SNR).

\textbf{Volume control} (VC) scales the magnitude of a speech signal, while not changing the semantics of the speech signal, but only its loudness. Since human perception of loudness is logarithmic, humans are not sensitive to linear changes in volume. In contrast, the model takes the audio as a sequence of sampling values that are linearly related to the volume.

\textbf{Fading} (FD), including fade in and fade out, adds a smooth transition to the beginning and end of an audio, which makes the audios sound more natural.
We apply five different fading shapes to the original signal, namely: linear, logarithmic, exponential, quarter sinusoidal and half sinusoidal.

\textbf{Time stretching} (TS) modifies the speed or duration of an audio signal without affecting its pitch. It is commonly used in audio signal processing to match the tempo of two songs, or to change the duration of an audio signal to fit a fixed length.

\textbf{Resampling} (RS) changes the sampling rate of an audio signal, a technique commonly employed in digital signal processing. This method converts signals from one sampling rate to another, ensuring compatibility between audio files, devices, or to decrease the size of an audio file.

\textbf{Time shifting} (SF) is a process to shift the audio in time domain, which can be used to delay or advance the audio signal by a certain amount of time. For human perception, regardless of how the audio is shifted, the auditory experience remains unchanged. However, for the model, the audio is a sequence of sampling values, the position of the sampling values do impact the model's performance. 

\textbf{Echoes adding} (EC) applies echoes to a speech signal by creating delayed and attenuated copies of the original signal and adding them together.
Adding echoes makes the speech sound more reverberant, depending on the delay and attenuation factors, while not affecting human recognition of speaker and content.

It is important to note that the manipulations described here are widely employed in audio processing. This simplifies the implementation and ensures that the resulting audio remains perceptually natural to humans.
\footnote{Our manipulated audio samples are accessible at the following URL: \url{https://sites.google.com/view/clad-demo/}.}
While prior research has considered pitch shift, time masking, and frequency masking~\cite{wang_deepsonar_2020,hojjati_self-supervised_2022}, we opt not to incorporate these manipulations into our study due to their potential to introduce perceptually unnatural audio artifacts and affect human recognition of speaker identity and content, contradicting the purpose of audio deepfake.

\subsection{Manipulation Attacks vs. Adversarial Attacks}

Manipulation attacks are similar to adversarial attacks in that both aim to evade detection systems and achieve malicious purposes. However, manipulation attacks offer three key advantages. First, manipulation attacks are simple and do not require professional knowledge of machine learning. Second, manipulation attacks are computationally efficient. Adversarial attacks is based on optimization, making the generation of successful adversarial examples expensive. In contrast, the manipulations used in our work can be applied quickly to audio deepfakes. The third advantage is that manipulation attacks do not need any prior knowledge of the detection model. Traditional adversarial attacks require full information of the target model, while black box adversarial attacks struggle with transferability issues~\cite{panariello_malafide_2023}. We believe these advantages make manipulation attacks more accessible to attackers, posing a serious threat to audio deepfake detection systems.

\begin{figure}[!t]
  \centering
  \subfigure[\cite{jung_aasist_2022} Original]{\includegraphics[width=0.142\textwidth]{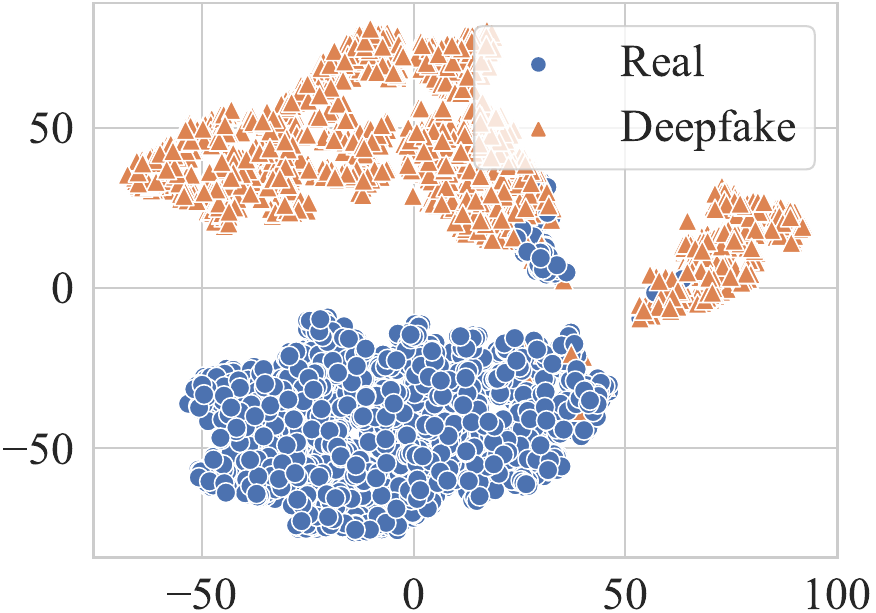}}\hspace{0.02\textwidth}
  \subfigure[\cite{tak_end--end_2021} Original]{\includegraphics[width=0.142\textwidth]{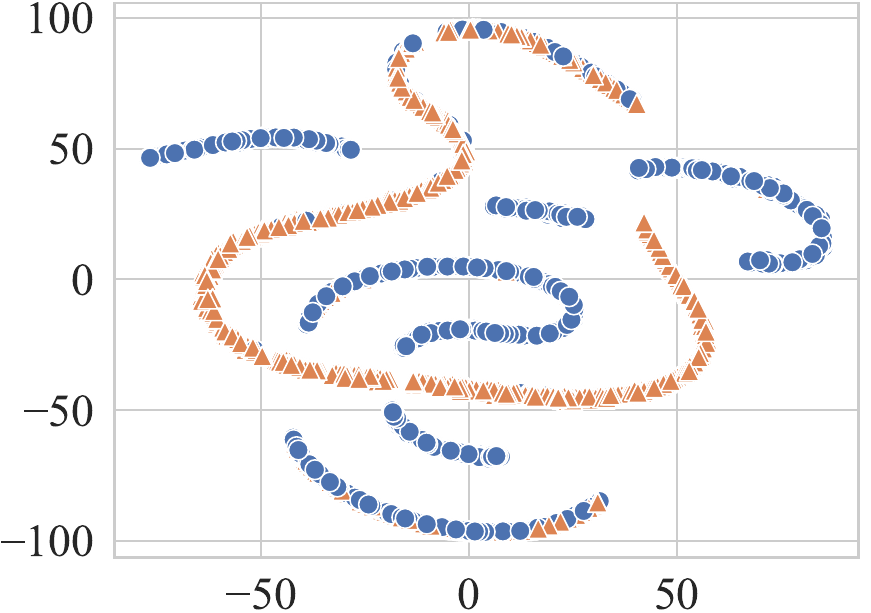}}\hspace{0.02\textwidth}
  \subfigure[\cite{hua_towards_2021} Original]{\includegraphics[width=0.142\textwidth]{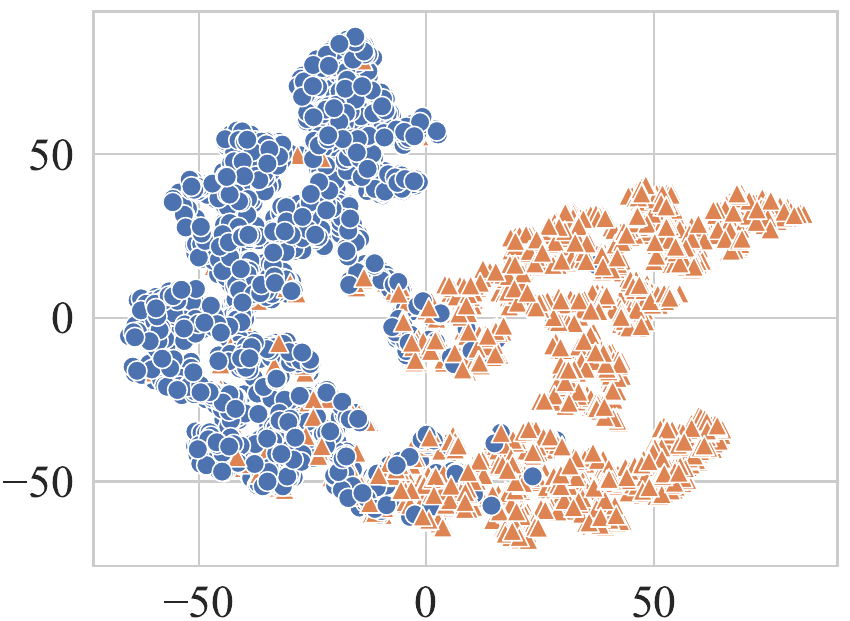}}\hspace{0.02\textwidth}
  \subfigure[\cite{jung_aasist_2022} Volume]{\includegraphics[width=0.142\textwidth]{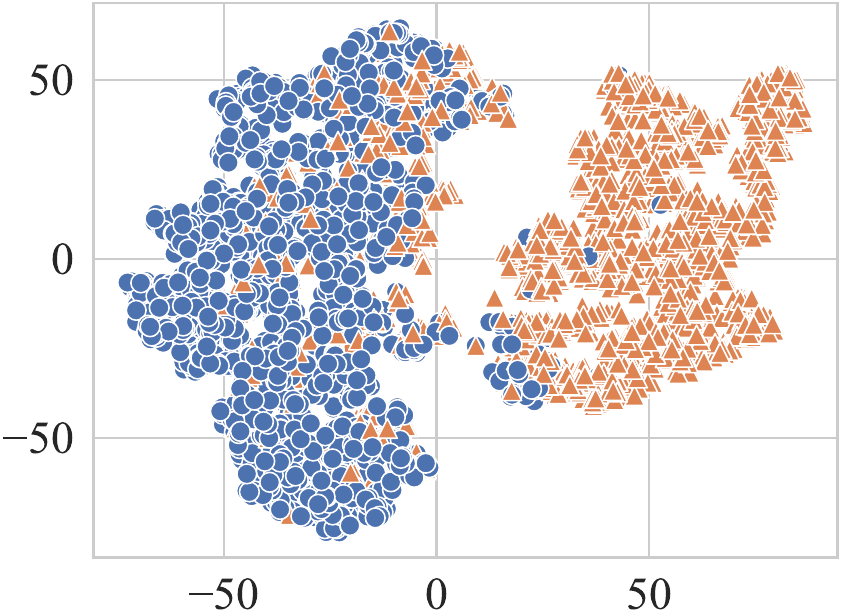}}\hspace{0.02\textwidth}
  \subfigure[\cite{tak_end--end_2021} Fade]{\includegraphics[width=0.142\textwidth]{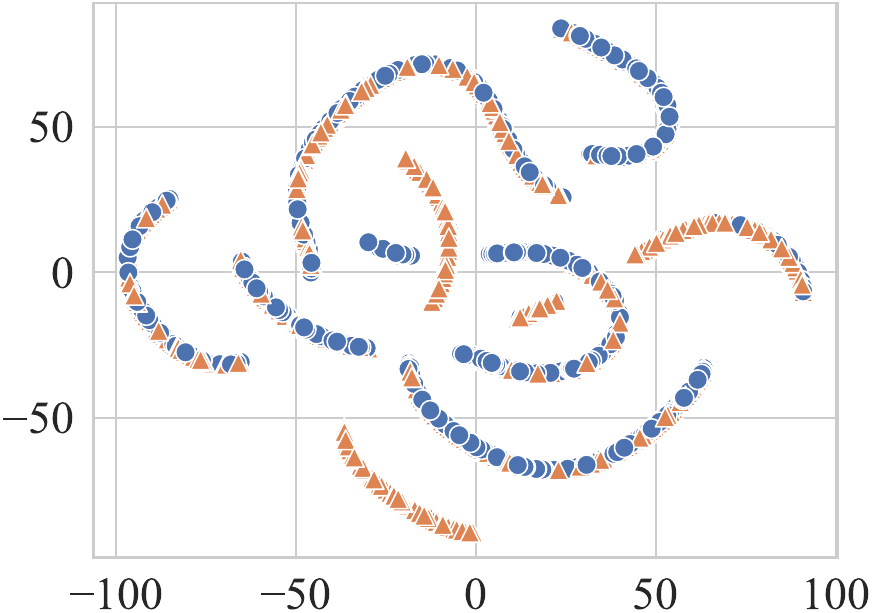}}\hspace{0.02\textwidth}
  \subfigure[\cite{hua_towards_2021} Noise]{\includegraphics[width=0.142\textwidth]{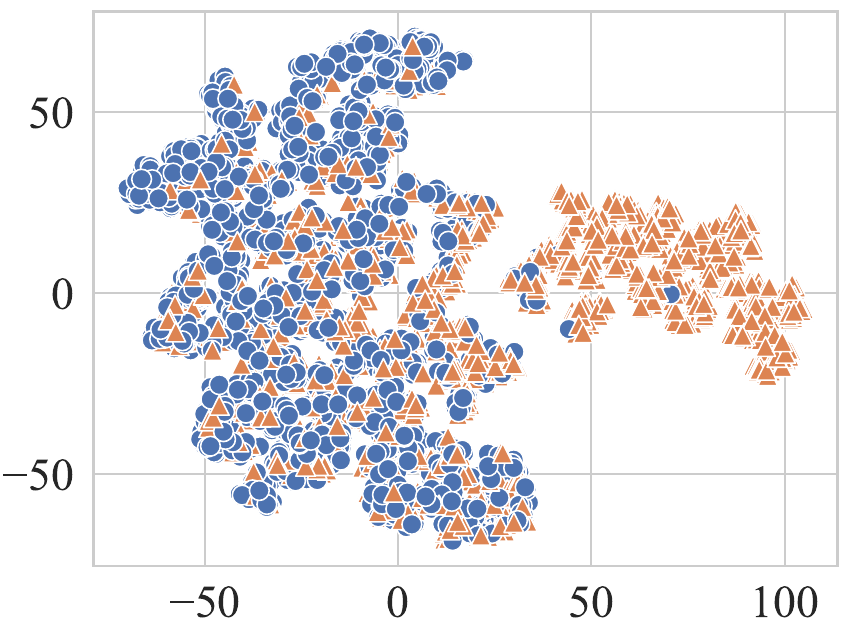}}\hspace{0.02\textwidth}
  \caption{Visualization of origianl vs. manipulated features extracted by widely adopted detection models via t-SNE. (a) and (d) are features extracted by AASIST, (b) and (e) are features extracted by RawNet2, (c) and (f) are features extracted by Res-TSSDNet.}\hspace{0.02\textwidth}
  \label{fig:manipulation_tsne_background}
\end{figure}

\section{Motivation}

To expose the impact of manipulation attacks, we employ three common techniques: volume control, fading, and noise injection as illustrations to manipulate the audio deepfake samples.
Fig.~\ref{fig:manipulation_tsne_background} visualizes the feature distribution extracted by three widely adopted models, AASIST~\cite{jung_aasist_2022}, RawNet2~\cite{tak_end--end_2021} and Res-TSSDNet~\cite{hua_towards_2021} via t-distributed stochastic neighbor embedding (t-SNE).
We can observe that compared with original distribution in Fig.~\ref{fig:manipulation_tsne_background}(a-c), the average distance between manipulated audio deepfakes and real samples is significantly reduced in Fig.~\ref{fig:manipulation_tsne_background}(d-f). Consequently, this leads to a notable degradation in the performance of the detection models when dealing with manipulated samples.

The underlying cause of this phenomenon is multifaceted. Firstly, existing detection methods have not accounted for the manipulations that could be potentially applied to the data. Secondly, these methods were predominantly trained using conventional supervised learning paradigms, thereby optimized to make predictions based on the specific examples they encountered during training. Manipulations cause the features extracted by these models changed significantly, resulting in the misclassification of the manipulated samples. This observation motivates us to design a robust model capable of withstanding potential manipulations. However, conventional supervised learning approaches often struggle to accommodate such variations due to their inherent limitations in adaptability.

In response to this challenge, we explore the potential of contrastive learning, a promising technique that empowers models to learn invariant and discriminative representations by distinguishing between similar and dissimilar examples based on their features. This enables models to capture the intrinsic characteristics of input audio, even in the presence of diverse manipulations. In contrast, conventional supervised learning approaches often struggle to handle such variations since they are tailored to specific instances and lack the inherent adaptability required. Motivated by these insights, we propose a contrastive learning-based detection model that learns more robust representations. 

Furthermore, we observed that that models trained solely on contrastive learning extract features with poor clustering properties, making it challenging to achieve satisfactory performance in the detection task. Given the observation that real samples are more similar to each other than audio deepfakes, we introduce length loss which clusters real audio representations by controlling length of the feature vector. While contrastive learning focuses on the directionality of features, length loss targets their magnitude, thereby not only avoiding interference with the contrastive learning process but also complementing it, boosting the model's performance.

\begin{figure}[!t]
  \centering
  \includegraphics[width=0.95\columnwidth]{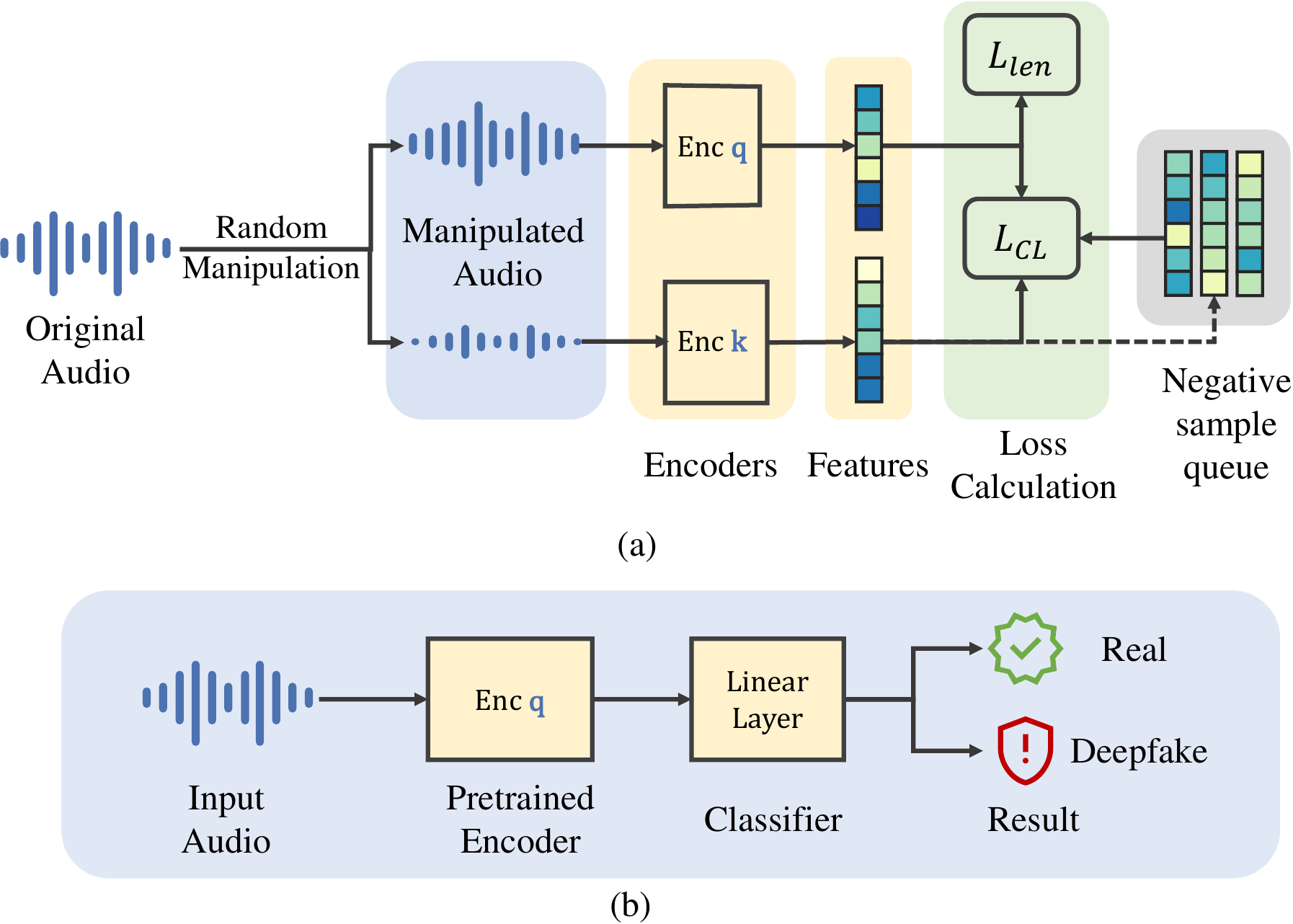}
  \caption{Overview of CLAD. (a) illustrates the pretraining stage. (b) illustrates the downstream training stage.}
  \label{fig:framework}
\end{figure}

\section{CLAD Design}

\subsection{Overview} 

CLAD is constructed based on the contrastive learning framework presented in ~\cite{he_momentum_2020}. We enhance the framework with additional audio augmentations and length loss. The CLAD model for audio deepfake detection task consists of two main components: an encoder and a linear classifier. The training of the CLAD model is conducted in two phases: pre-training and downstream training, as s illustrated in Fig.~\ref{fig:framework}.

During pre-training, our objective is to train a robust encoder that can produce representative features. Firstly, we randomly apply different manipulations to the input training data, resulting in two manipulated samples. We then feed the manipulated samples into the encoders to obtain the corresponding features.  In the context of contrastive learning, samples augmented from the same audio are deemed positive pairs, while those from different audios are considered negative pairs. The encoder is trained to produce similar features for positive pairs and dissimilar features for negative pairs. This training paradigm facilitates the model in learning to generate consistent features for samples, even when subjected to different manipulations, while still maintaining discriminative features across different audios. To enhance the contrastive training, we employ a queue to store previous negative samples, as suggested in ~\cite{he_momentum_2020}. The contrastive loss is computed using the features output by the encoders and negative samples stored in the queue. Additionally, we introduce length loss to aggregate features for real samples, thereby augmenting downstream performance. Further details regarding the contrastive loss and length loss are elaborated in following sections. The final loss function for pre-training is expressed as the weighted sum of the contrastive loss and length loss, as shown in Eq.~\ref{eq:pretrain_loss}.

\begin{equation}
  \label{eq:pretrain_loss}
  \mathcal{L}_{pretrain} = \mathcal{L}_{CL} + \lambda\mathcal{L}_{len}
\end{equation}

In the downstream training stage, we utilize the pre-trained encoder and add a linear classifier on top of it. We then train the model on the audio deepfake detection task using the labeled dataset to obtain the final detector.

Note that we do not specify the encoder architecture in CLAD.
Instead, most deep learning-based end-to-end detection methods can be used as the encoder by removing the classification head.
This makes CLAD a plug-and-play solution that can be easily integrated with existing detection methods to improve their robustness.

\subsection{Contrastive Learning}

Contrastive learning aims to learn consistent and discriminative representations by training a model to map similar instances closer together and dissimilar instances farther apart in a latent feature space. We employ contrastive learning to train the encoder to produce consistent representations under various manipualtions, enhancing the model's resistance against manipulation attacks.

While prior contrastive learning methods for audio~\cite{saeed_contrastive_2021,guan_anomalous_2023} primarily relied on the SimCLR framework~\cite{chen_simple_2020}, we opt for MoCo\cite{he_momentum_2020} because it allows for the use of a larger number of negative samples through the maintenance of a negative sample queue.
This framework employs two different encoders, named query encoder and key encoder, which are the Enc q and Enc k shown in Fig.~\ref{fig:framework}.  The encoders have identical architecture, and the parameters of the key encoder are smoothly updated by using the parameters of the query encoder.
\footnote{The design of momentum update and negative sample queue aims to increase the negative sample size and stabilize the training process, thereby enhancing the model performance. We refer readers to \cite{he_momentum_2020} for more details.}
They take the different augmentations of input and produce corresponding features, $q$ and $k$. The augmentations include all the manipulations discussed in Sec.~\ref{sec:manipulation_attacks} and are randomly applied during training. We train the encoder using contrastive loss to ensure that the generated feature $k$ is as close as possible to feature $q$ compared to the negative sample features in the queue. This enables the encoder to distinguish between different samples under manipulations. The contrastive loss is defined in Eq.~\ref{eq:contrastive_loss}.
\begin{equation}
  \label{eq:contrastive_loss}
  \mathcal{L}_{CL} = -\frac{1}{N}\sum_{i=1}^N \log \frac{\exp (q_i^Tk_i^+/\tau)}{\sum_{j=1}^K \exp (q_i^Tk_j/\tau)}
\end{equation}
where $N$ is the batch size, $K$ is the size of the negative sample queue, $\tau$ is the temperature parameter, and $q_i$ and $k_i^+$ are the feature q and feature k of the $i$-th sample in the batch and $k_j$ is the feature k of the $j$-th sample in the negative sample queue. After the contrastive loss is computed, we update the parameters of the query encoder with gradient descent. The parameters of the key encoder are updated by using the momentum update rule formulated in Eq.~\ref{eq:momentum_update}.

\begin{equation}
  \label{eq:momentum_update}
  \theta_k = \mu\theta_k + (1-\mu)\theta_q
\end{equation}
where $\theta_k$ and $\theta_q$ are the parameters of the key encoder and the query encoder, respectively, and $\mu$ is the momentum parameter.
Finally, we enqueue the feature $k$ in the batch to the negative sample queue and dequeue the earliest features in the queue to update the queue.

\begin{figure}[!t]
  \centering
  \includegraphics[width=0.9\columnwidth]{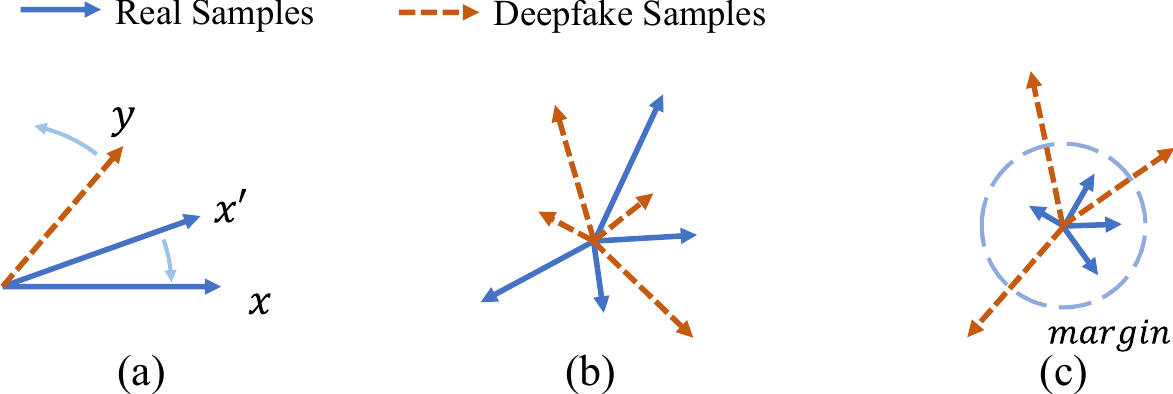}
  \caption{Illustration of the motivation of length loss. (a) illustrates the training of contrastive loss. (b) illustrates the features extracted by contrastive loss trained encoder. (c) illustrates the features extracted by length loss and contrastive loss trained encoder.}
  \label{fig:len_loss_illustration}
\end{figure}

\subsection{Length Loss}

The contrastive loss we used is based on the cosine similarity. However, by solely relying on the cosine similarity for calculating the contrastive loss, the model only learns to minimize the angles between feature vectors corresponding to augmented audios from the same source while increasing the angles for different audios, as depicted in Fig.~\ref{fig:len_loss_illustration}~(a). After such training, we observed that the features extracted by the model remain scattered and exhibit poor clustering, as illustrated in Fig.~\ref{fig:len_loss_illustration}~(b). In the context of audio deepfake detection, real speech samples tend to exhibit higher similarity to one another, whereas synthetic speech, generated using different algorithms, often poses challenges for clustering~\cite{zhang_one_class_2021}. Building upon this observation, we propose clustering the feature vectors of real audio to enhance performance. To accomplish this, we introduce length loss that leverages the length information, which is not utilized in the cosine similarity loss calculation. Specifically, we decrease the length of feature vectors for real speech and increase it for audio deepfakes, leading to improved clustering of feature vectors for real speech, as demonstrated in Fig.~\ref{fig:len_loss_illustration}~(c). The length loss is defined mathematically as Eq.~\ref{eq:len_loss}.

\begin{equation}
  \label{eq:len_loss}
  \mathcal{L}_{len} = \frac{1}{N}\sum_{i=1}^N y_i w  \|q_i\|_2 + (1-y_i)  \max (margin-\|q_i\|_2,0)
\end{equation}
where $N$ is the batch size, $q_i$ denotes the feature of the $i$-th sample in the batch extracted by encoder $Enc\,q$, $y_i$ is the label of the sample, 
$w$ is the weight factor to assign different levels of importance to the different classes, since the audio deepfake detection dataset might be highly unbalanced. $margin$ is the margin used to control the degree of separation between the real audios and audio deepfakes.

In summary, length loss is designed to encourage the encoder to generate shorter feature vectors for real audios and longer ones for audio deepfakes. This is motivated by the observation that real speech samples generally exhibit more similarity compared to audio deepfakes generated by various algorithms. Besides, when integrated with the contrastive loss, which focuses on the directionality of feature vectors, the length loss provides an additional dimension of discriminative information. This combination of loss functions allows the model to leverage both magnitude and direction of feature vectors, resulting in improved performance in detection. 

\subsection{Downstream Training}

After training the encoder with both the contrastive and length loss, we combine the encoder Enc q with a linear layer and train it by minimizing the cross-entropy loss for the downstream task. The linear classifier is a one-layer fully connected neural network that maps the encoder output to the classification score. We define the loss function as follows:

\begin{equation} 
  \label{eq:downstream_loss}
  \mathcal{L}_{downstream} = -\frac{1}{N}\sum_{i=1}^N  [y_i\log(p_i) + (1-y_i)\log(1-p_i)]
\end{equation}
where $N$ is the batch size, $y_i$ is the label, $p_i$ is the predicted probability of being real for the $i$-th sample in the batch.

\begin{table*}[!t]
  \fontsize{6.2pt}{7.4pt}\selectfont
  \caption{The FARs (\%) and F1 scores (\%) of different audio deepfake detection methods under manipulation attacks.}
  \label{tab:comparison_under_attacks}
  \begin{threeparttable}
    \begin{tabular}{l|c|l|ll|lll|lllllll|llll}
      \hline

      Models  & Types & None                                     & \multicolumn{2}{c|}{Volume Control}     & \multicolumn{3}{c|}{White Noise} & \multicolumn{7}{c|}{Environmental Noise\tnote{1}} & \multicolumn{4}{c}{Time Stretch}                                                                                                  \\
      ~       & ~     & ~                                        & 0.5                             & 0.1                                   & 15dB                                              & 20dB                             & 25dB  & WD    & FS    & BR    & CO    & RA    & CT    & SN     & 1.1   & 1.05  & 0.95  & 0.9   \\
      \hline
      RawNet2 & FAR & 4.60  & 2.78  & \textbf{36.62} & \textbf{18.00} & 10.09 & 7.44  & 4.02  & 2.95  & 3.74  & 4.30  & 8.68  & 7.65  & 3.37  & 0.28  & 0.44  & 0.33  & 0.16  \\
      ~       & F1  & 81.08 & 86.90 & \textbf{37.16} & \textbf{54.24} & 67.42 & 73.38 & 82.86 & 86.34 & 83.74 & 81.99 & 70.46 & 72.88 & 84.94 & 96.43 & 95.78 & 96.24 & 96.95 \\
      \hline
      AASIST  & FAR & 0.83  & 1.49  & \textbf{9.12}  & 0.07  & 0.13  & 0.40  & 0.08  & 0.26  & 0.17  & 0.10  & 0.10  & 0.11  & 0.13  & 0.16  & 0.08  & 0.04  & 0.03  \\
      ~       & F1  & 96.11 & 93.51 & \textbf{71.25} & 99.30 & 99.02 & 97.89 & 99.26 & 98.45 & 98.84 & 99.15 & 99.14 & 99.12 & 99.04 & 98.89 & 99.26 & 99.43 & 99.45 \\
      \hline
      Res-TSSDNet & FAR & 1.63  & 2.19  & \textbf{10.11} & \textbf{51.28} & \textbf{39.23} & \textbf{29.80} & \textbf{36.68} & 12.55 & 3.91  & 1.68  & \textbf{40.37} & 12.70 & 2.84  & 0.19  & 0.20  & 0.00  & 0.00  \\
      ~       & F1    & 92.57 & 90.50 & \textbf{68.74} & \textbf{30.56} & \textbf{36.49} & \textbf{43.03} & \textbf{38.05} & 64.01 & 84.69 & 92.39 & \textbf{35.84} & 63.74 & 88.19 & 98.34 & 98.31 & 99.17 & 99.17 \\
      \hline
      SAMO & FAR   & 1.09   & 3.11   & \textbf{7.67}  & 0.22     & 0.77     & 1.40     & 0.38  & 0.29  & 0.58  & 0.12  & 0.58  & 0.32  & 0.46   & 0.00  & 0.01  & 0.00  & 0.00  \\
      ~       & F1    & 94.94  & 87.56  & \textbf{74.50} & 98.51    & 96.22    & 93.74    & 97.83 & 98.19 & 97.01 & 98.94 & 96.98 & 98.09 & 97.48  & 99.44 & 99.43 & 99.45 & 99.45 \\
      \hline
      CLAD    & FAR & 1.11  & 0.70  & 0.06  & 0.11  & 0.51  & 0.82  & 1.39  & 1.17  & 0.68  & 0.20  & 0.23  & 1.15  & 1.01  & 0.07  & 0.12  & 0.06  & 0.03  \\
      ~       & F1  & 94.82 & 96.48 & 99.17 & 98.95 & 97.26 & 96.01 & 93.74 & 94.59 & 96.56 & 98.60 & 98.43 & 94.68 & 95.25 & 99.13 & 98.91 & 99.18 & 99.32  \\
      \hline
    \end{tabular}

    \vspace{1.5em}  

    \begin{tabular}{l|c|lll|lll|lllllll|llll}
      \hline
      Models  & Types & \multicolumn{3}{c|}{Add Echoes\tnote{2}} & \multicolumn{3}{c|}{Time Shift} & \multicolumn{7}{c|}{Fade\tnote{3}} & \multicolumn{4}{c}{Resample}                                                                                                          \\
      ~       & ~     & 1k/ .2                                   & 1k/ .5                          & 2k/ .5                                & 1.6k                                              & 16k                              & 32k   & .5/ L & .3/ L & .1/ L & .5/ E & .5/ Q & .5/ H & .5/ Lo & -1k   & -0.5k & +0.5k & +1k   \\
      \hline
      RawNet2 & FAR & 3.55   & 1.48   & 2.71   & 4.08  & 3.99  & 2.14  & 15.57 & 7.61  & 4.79  & \textbf{27.09} & 7.88  & \textbf{30.42} & 4.58   & 3.82  & 3.92  & 3.87  & 3.67  \\
      ~       & F1  & 84.34  & 91.62  & 87.14  & 82.66 & 82.92 & 89.16 & 57.71 & 72.97 & 80.50 & \textbf{44.30} & 72.31 & \textbf{41.52} & 81.14  & 83.48 & 83.15 & 83.33 & 83.95 \\
      \hline
      AASIST  & FAR & 0.23   & 0.07   & 0.10   & 0.60  & 0.75  & 0.54  & 9.63  & 3.60  & 1.05  & \textbf{23.53} & 3.83  & \textbf{31.22} & 1.23   & 0.59  & 0.60  & 0.97  & 1.25  \\
      ~       & F1  & 98.59  & 99.29  & 99.16  & 97.05 & 96.45 & 97.31 & 70.13 & 86.08 & 95.24 & \textbf{49.15} & 85.32 & \textbf{42.17} & 94.51  & 97.07 & 97.03 & 95.55 & 94.43 \\
      \hline
      Res-TSSDNet & FAR & 2.18   & 1.29   & 1.61   & 2.38  & 2.25  & 2.15  & 2.80  & 1.68  & 1.17  & 4.91  & 1.83  & \textbf{11.55} & 1.47   & 0.00  & 0.00  & 1.78  & 2.08  \\
      ~       & F1    & 90.53  & 93.88  & 92.66  & 89.83 & 90.29 & 90.64 & 88.34 & 92.39 & 94.35 & 81.62 & 91.82 & \textbf{65.87} & 93.17  & 99.18 & 99.18 & 92.01 & 90.89 \\
      \hline
      SAMO  & FAR & 0.15   & 0.04   & 0.07  & 0.69     & 1.89     & 1.58     & \textbf{17.65} & 10.34 & 2.16  & \textbf{25.98} & 11.07 & \textbf{28.24} & 4.53   & 0.38  & 0.43  & 1.19  & 1.55  \\
      ~     & F1  & 98.78  & 99.28  & 99.17 & 96.56    & 91.89    & 93.05    & \textbf{56.17} & 68.51 & 90.89 & \textbf{46.60} & 67.05 & \textbf{44.53} & 83.01  & 97.84 & 97.64 & 94.55 & 93.16 \\
      \hline
      CLAD    & FAR & 0.21   & 0.03   & 0.07   & 0.85  & 0.93  & 0.89  & 0.64  & 0.95  & 1.06  & 0.50  & 0.85  & 0.80  & 0.93   & 0.65  & 0.81  & 1.37  & 1.63  \\
      ~       & F1  & 98.55  & 99.32  & 99.15  & 95.88 & 95.57 & 95.74 & 96.73 & 95.50 & 95.05 & 97.32 & 95.88 & 96.07 & 95.57  & 96.68 & 96.03 & 93.81 & 92.83  \\
      \hline
    \end{tabular}
    \begin{tablenotes}
      \item[1] The abbreviations WD, FS, BR, CO, RA, CT, SN stand for wind, footsteps, breathing, coughing, rain, clock tick and sneezing respectively.
      \item[2] 1k/ .2 means that the delay is 1,000 samples and the attenuation factor is 0.2, and the rest is the same.
      \item[3] .5/ L means the ratio is set to 0.5, and using linear fade shape. L, E, Q, H, Lo denote linear, exponential, quarter sinusoidal, half sinusoidal, logarithmic fade shapes respectively.
    \end{tablenotes}
  \end{threeparttable}

\end{table*}

\section{Experimental Results}

\noindent The objective of our evaluation is to answer the following Research Questions(RQs):

\begin{itemize}[leftmargin=*]
  \item RQ1: How do manipulation attacks degrade the performance of widely adopted audio deepfake detection systems? 
  \item RQ2: How effective is the proposed CLAD model in detecting audio deepfakes under manipulation attacks?
  \item RQ3: What is the impact on the performance of the CLAD model when contrastive learning or length loss is removed?
  \item RQ4: How does the performance of the CLAD model change when facing unknown manipulations?
\end{itemize}
\subsection{Experimental Settings}  
\label{sec:implementation_details}
\textbf{Baseline Models}.
We select RawNet2~\cite{tak_end--end_2021}, AASIST~\cite{jung_aasist_2022}, Res-TSSDNet~\cite{hua_towards_2021}, and SAMO~\cite{ding_samo_2023} as baseline models for audio deepfake detection. They are the most widely adopted and influential models in this field, selected as the baseline models for many challenges~\cite{yamagishi_asvspoof_2021, jung_sasv_2022, yi_add_2022}, and have been studied in recent works~\cite{ba_transferring_2023, muller_does_2022, zhang_compressed_2023}.
Furthermore, these models are accompanied by publicly available official implementations and pretrained model weights. 
To assess their performance against manipulation attacks, we utilize the pretrained model weights provided by the authors instead of training the models from scratch. For models with multiple versions, we select the one with the best performance reported.

\textbf{Dataset}.
We used the Logical Access (LA) part of the ASVspoof 2019 dataset, which is commonly utilized by audio deepfake detection methods ~\cite{tak_end--end_2021,jung_aasist_2022,wen_multi_path_2022,wang_investigating_2022,conti_deepfake_2022, doan_bts_2023}, including the baseline models.
We followed baseline methods and trained our models exclusively on the training set of the ASVspoof 2019 LA to ensure a fair comparison.
For evaluation, we used only the evaluation set of ASVspoof 2019 LA, excluding the other datasets such as ASVspoof 2021 LA and ASVspoof 2021 DF.
This is due to the subpar performance of the pretrained baseline models on these datasets, making it challenging to demonstrate the impact of manipulation attacks.

\textbf{Metrics}.
The evaluation metrics utilized in our study comprise the False Acceptance Rate (FAR), False Rejection Rate (FRR), F1 score, and Equal Error Rate (EER). FAR represents the proportion of deepfakes falsely classified as real audio, whereas FRR represents the opposite. Notably, FAR serves as a measure of the success rate of manipulation attacks and is therefore adopted as the primary metric for our evaluation.
F1 score is computed as the harmonic mean of precision and recall. The EER corresponds to the point on the Detection Error Tradeoff (DET) curve where the FAR equals the FRR. Notably, the calculation of FAR, FRR, and F1 score necessitates a threshold, and for most evaluations, we adopt the threshold that yields the EER on the original data.
Thus in our evaluation, FAR is also the EER on the original data.

\textbf{Model hyperparameters.}
We take AASIST~\cite{jung_aasist_2022} model without the final fully connected layer as the encoder of our model to demonstrate the improvement of CLAD. The input audio is repeated or clipped to make the duration is fixed to 64600 samples.
For pretraining stage, the model is trained with Adam optimizer using a learning rate of 0.0005, a weight decay of 0.0001, 150 epochs and a mini-batch size of 24. We also use the cosine annealing learning rate decay following the strategy of MoCo and AASIST. 
The queue size for contrastive learning is reduced to 6144, since large queue size will cause the queue to be filled with the same samples. 
The temperature $\tau$ and the momentum $\mu$ are set to 0.07 and 0.999, respectively. Concerning Length loss, we set the margin $margin$ to 4 and the weight $w$ to 9. The weight of the length loss $\alpha$ is set to 2. The pretrain epochs is selected as 150 empirically.
For downstream training stage, we train the model with Adam optimizer using a learning rate of 0.001, a weight decay of 0.0001, 10 epochs and a mini-batch size of 16.

\textbf{Manipulation parameters}
To evaluate the impact of white noise, we varied the signal-to-noise ratio (SNR) with values of 15, 20, and 25dB.
Regarding environmental noise, we added 7 types of noise from the ESC-50 dataset ~\cite{piczak2015dataset} at a constant SNR of 20dB.
For volume control, we examined factors of 0.5 and 0.1.
To investigate the effect of fading, we used a linear fade shape and varied the fade ratio with values of 0.1, 0.3, and 0.5.
Additionally, we examined the influence of five fade shapes, each employing a consistent fade ratio of 0.5.
To evaluate the impact of time stretching, we used an FFT length of 128 and varied the time stretching factor with values of 0.9, 0.95, 1.05, and 1.1.
For resampling, we evaluated target resampling rates of 15,000, 15,500, 16,500, and 17,000 Hz, which correspond to -1,000, -500, +500, and +1,000 offsets to the original sampling rate of 16,000 Hz.
We examined time shifting by considering shift lengths of 1,600, 16,000, and 32,000.
To add echoes, we studied two parameters: delay and attenuation factor. We set the delay to 1,000 or 2,000 and the attenuation to 0.2 or 0.5.
Additional information about the manipulation parameters is provided in the appendix.

\begin{table}[t!]
  \caption{The FARs (\%) of different audio deepfake detection methods under combination of manipulation attacks.}
  \label{tab:combination_manipulations}
  \centering
  \scriptsize  
  \renewcommand{\arraystretch}{1.2} 
  \subfigure[RawNet2]{
  \centering
  \begin{tabular}{p{45pt}p{16pt}p{16pt}p{16pt}p{16pt}p{16pt}p{16pt}}
    \rowcolor[HTML]{ECECEC} 
    & VC                            & WN                            & EN                            & TS                           & FD                            & RS                            \\
 VC 0.1 & \cellcolor[HTML]{FBBEC1}36.69 & \cellcolor[HTML]{F8696B}86.47 & \cellcolor[HTML]{FA9799}59.9  & \cellcolor[HTML]{FCEFF2}7.87 & \cellcolor[HTML]{FBC3C6}33.57 & \cellcolor[HTML]{FBC3C5}34.07 \\
 WN 15dB & \cellcolor[HTML]{F96A6C}86.45 & \cellcolor[HTML]{FCDEE1}18.02 & \cellcolor[HTML]{FBD7DA}21.77 & \cellcolor[HTML]{FCF4F7}5.09 & \cellcolor[HTML]{F97F81}73.86 & \cellcolor[HTML]{FCE9EC}11.46 \\
 EN Wind & \cellcolor[HTML]{FA9799}59.91 & \cellcolor[HTML]{FBD7DA}21.90  & \cellcolor[HTML]{FCEEF1}8.68  & \cellcolor[HTML]{FCFAFD}1.30  & \cellcolor[HTML]{FA999B}58.67 & \cellcolor[HTML]{FCF0F3}7.14  \\
 TS 0.9 & \cellcolor[HTML]{FCEFF2}7.89  & \cellcolor[HTML]{FCF1F4}6.53  & \cellcolor[HTML]{FCFAFD}1.67  & \cellcolor[HTML]{FCFCFF}0.16 & \cellcolor[HTML]{FCFAFD}1.31  & \cellcolor[HTML]{FCFCFF}0.09  \\
 FD .5/ H & \cellcolor[HTML]{FBC3C6}33.57 & \cellcolor[HTML]{F98385}71.7  & \cellcolor[HTML]{FA9DA0}56.02 & \cellcolor[HTML]{FCFBFE}1.01 & \cellcolor[HTML]{FBC9CB}30.43 & \cellcolor[HTML]{FBCED1}27.35 \\
 RS +1k & \cellcolor[HTML]{FBC3C5}34.07 & \cellcolor[HTML]{FCDADD}20.03 & \cellcolor[HTML]{FCEDF0}9.14  & \cellcolor[HTML]{FCFCFF}0.09 & \cellcolor[HTML]{FBD0D3}26.02 & \cellcolor[HTML]{FCF6F9}3.67 
 \end{tabular}%
  }%

  \subfigure[AASIST]{
  \centering
  \begin{tabular}{p{45pt}p{16pt}p{16pt}p{16pt}p{16pt}p{16pt}p{16pt}}
    \rowcolor[HTML]{ECECEC} 
    & VC                            & WN                            & EN                            & TS                           & FD                            & RS                            \\
 VC 0.1 & \cellcolor[HTML]{FCEDF0}9.15  & \cellcolor[HTML]{FCE7EA}12.64 & \cellcolor[HTML]{FCEAED}10.61 & \cellcolor[HTML]{FCFCFF}0.45 & \cellcolor[HTML]{FCD9DC}20.71 & \cellcolor[HTML]{FCECEF}9.83  \\
 WN 15dB & \cellcolor[HTML]{FCE7EA}12.71 & \cellcolor[HTML]{FCFCFF}0.07  & \cellcolor[HTML]{FCFCFF}0.06  & \cellcolor[HTML]{FCFCFF}0.03 & \cellcolor[HTML]{FCE2E5}15.51 & \cellcolor[HTML]{FCFCFF}0.14  \\
 EN Wind & \cellcolor[HTML]{FCEAED}10.61 & \cellcolor[HTML]{FCFCFF}0.06  & \cellcolor[HTML]{FCFCFF}0.10   & \cellcolor[HTML]{FCFCFF}0.03 & \cellcolor[HTML]{FBCFD2}26.74 & \cellcolor[HTML]{FCFCFF}0.24  \\
 TS 0.9 & \cellcolor[HTML]{FCFCFF}0.45  & \cellcolor[HTML]{FCFCFF}0.03  & \cellcolor[HTML]{FCFCFF}0.03  & \cellcolor[HTML]{FCFCFF}0.03 & \cellcolor[HTML]{FCFCFF}0.31  & \cellcolor[HTML]{FCFCFF}0.03  \\
 FD .5/ H & \cellcolor[HTML]{FCD9DC}20.71 & \cellcolor[HTML]{FCFBFE}0.96  & \cellcolor[HTML]{FCF7FA}3.28  & \cellcolor[HTML]{FCFBFE}0.59 & \cellcolor[HTML]{FBC7CA}31.23 & \cellcolor[HTML]{FBC9CB}30.57 \\
 RS +1k & \cellcolor[HTML]{FCECEF}9.83  & \cellcolor[HTML]{FCFCFF}0.16  & \cellcolor[HTML]{FCFCFF}0.23  & \cellcolor[HTML]{FCFCFF}0.07 & \cellcolor[HTML]{FBC6C8}32.14 & \cellcolor[HTML]{FCFAFD}1.25 
 \end{tabular}%
}%

\subfigure[Res-TSSDNet]{
  \centering
  \begin{tabular}{p{45pt}p{16pt}p{16pt}p{16pt}p{16pt}p{16pt}p{16pt}}
    \rowcolor[HTML]{ECECEC} 
    & VC                            & WN                            & EN                            & TS                           & FD                            & RS                            \\
 VC 0.1 & \cellcolor[HTML]{FCEBEE}10.11 & \cellcolor[HTML]{F97F81}73.62 & \cellcolor[HTML]{F98587}70.24 & \cellcolor[HTML]{FCFBFE}0.78 & \cellcolor[HTML]{FCF2F5}6.18  & \cellcolor[HTML]{FCEAED}10.61 \\
 WN 15dB & \cellcolor[HTML]{F97F81}73.63 & \cellcolor[HTML]{FAA5A8}51.28 & \cellcolor[HTML]{FAA0A2}54.53 & \cellcolor[HTML]{FCFCFF}0.00    & \cellcolor[HTML]{FAA9AB}49.17 & \cellcolor[HTML]{FAA3A6}52.57 \\
 EN Wind & \cellcolor[HTML]{F98587}70.24 & \cellcolor[HTML]{FAA0A2}54.43 & \cellcolor[HTML]{FBB8BA}40.37 & \cellcolor[HTML]{FCFCFF}0.00    & \cellcolor[HTML]{FBB3B6}43.00    & \cellcolor[HTML]{FBB5B7}42.12 \\
 TS 0.9 & \cellcolor[HTML]{FCFBFE}0.78  & \cellcolor[HTML]{FCF0F3}7.45  & \cellcolor[HTML]{FCF8FB}2.62  & \cellcolor[HTML]{FCFCFF}0.00    & \cellcolor[HTML]{FCFCFF}0.13  & \cellcolor[HTML]{FCFCFF}0.00     \\
 FD .5/ H & \cellcolor[HTML]{FCF2F5}6.18  & \cellcolor[HTML]{FA9D9F}56.3  & \cellcolor[HTML]{FAA6A9}50.82 & \cellcolor[HTML]{FCFCFF}0.39 & \cellcolor[HTML]{FCE9EC}11.55 & \cellcolor[HTML]{FCE0E3}16.77 \\
 RS +1k & \cellcolor[HTML]{FCEAED}10.61 & \cellcolor[HTML]{FCFBFE}0.96  & \cellcolor[HTML]{FCF7FA}3.28  & \cellcolor[HTML]{FCFBFE}0.59 & \cellcolor[HTML]{FBC9CB}30.57 & \cellcolor[HTML]{FCF9FC}2.08 
 \end{tabular}%
}%

\subfigure[SAMO]{
  \centering
  \begin{tabular}{p{45pt}p{16pt}p{16pt}p{16pt}p{16pt}p{16pt}p{16pt}}
    \rowcolor[HTML]{ECECEC} 
    & VC                           & WN                           & EN                           & TS                           & FD                           & RS                           \\
 VC 0.1 & \cellcolor[HTML]{FCEFF2}7.67  & \cellcolor[HTML]{FCEBED}10.54 & \cellcolor[HTML]{FCEDF0}9.00  & \cellcolor[HTML]{FCFCFF}0.00 & \cellcolor[HTML]{FCF3F6}5.36  & \cellcolor[HTML]{FCEFF2}8.08  \\
 WN 15dB & \cellcolor[HTML]{FCEBEE}10.46 & \cellcolor[HTML]{FCFCFF}0.22  & \cellcolor[HTML]{FCFCFF}0.15  & \cellcolor[HTML]{FCFCFF}0.00 & \cellcolor[HTML]{FCD8DB}21.48 & \cellcolor[HTML]{FCFCFF}0.27  \\
 EN Wind & \cellcolor[HTML]{FCEDF0}9.00  & \cellcolor[HTML]{FCFCFF}0.15  & \cellcolor[HTML]{FCFCFF}0.38  & \cellcolor[HTML]{FCFCFF}0.00 & \cellcolor[HTML]{FBCFD2}26.48 & \cellcolor[HTML]{FCFBFE}0.79  \\
 TS 0.9 & \cellcolor[HTML]{FCFCFF}0.00  & \cellcolor[HTML]{FCFCFF}0.01  & \cellcolor[HTML]{FCFCFF}0.01  & \cellcolor[HTML]{FCFCFF}0.00 & \cellcolor[HTML]{FCFCFF}0.03  & \cellcolor[HTML]{FCFCFF}0.00  \\
 FD .5/ H & \cellcolor[HTML]{FCF3F6}5.36  & \cellcolor[HTML]{FCE0E2}16.96 & \cellcolor[HTML]{FBCFD2}26.64 & \cellcolor[HTML]{FCFCFF}0.06 & \cellcolor[HTML]{FBCCCF}28.24 & \cellcolor[HTML]{FBCACD}29.79 \\
 RS +1k & \cellcolor[HTML]{FCEFF2}8.08  & \cellcolor[HTML]{FCFCFF}0.27  & \cellcolor[HTML]{FCFBFE}0.74  & \cellcolor[HTML]{FCFCFF}0.00 & \cellcolor[HTML]{FBC9CC}30.11 & \cellcolor[HTML]{FCFAFD}1.55 
 \end{tabular}%
}%

\subfigure[CLAD]{
  \centering
  \begin{tabular}{p{45pt}p{16pt}p{16pt}p{16pt}p{16pt}p{16pt}p{16pt}}
    \rowcolor[HTML]{ECECEC} 
    & VC                           & WN                           & EN                           & TS                           & FD                           & RS                           \\
 VC 0.1 & \cellcolor[HTML]{FCFCFF}0.06\phantom{0} & \cellcolor[HTML]{FCFCFF}0.03\phantom{0} & \cellcolor[HTML]{FCFCFF}0.04\phantom{0} & \cellcolor[HTML]{FCFCFF}0.00\phantom{0}    & \cellcolor[HTML]{FCFCFF}0.00\phantom{0}    & \cellcolor[HTML]{FCFCFF}0.08\phantom{0} \\
 WN 15dB & \cellcolor[HTML]{FCFCFF}0.02 & \cellcolor[HTML]{FCFCFF}0.12 & \cellcolor[HTML]{FCFCFF}0.05 & \cellcolor[HTML]{FCFCFF}0.01 & \cellcolor[HTML]{FCFCFF}0.33 & \cellcolor[HTML]{FCFCFF}0.38 \\
 EN Wind & \cellcolor[HTML]{FCFCFF}0.00    & \cellcolor[HTML]{FCFCFF}0.00    & \cellcolor[HTML]{FCFCFF}0.23 & \cellcolor[HTML]{FCFCFF}0.00    & \cellcolor[HTML]{FCFCFF}0.19 & \cellcolor[HTML]{FCFCFF}0.31 \\
 TS 0.9 & \cellcolor[HTML]{FCFCFF}0.00    & \cellcolor[HTML]{FCFCFF}0.00    & \cellcolor[HTML]{FCFCFF}0.01 & \cellcolor[HTML]{FCFCFF}0.03 & \cellcolor[HTML]{FCFCFF}0.10  & \cellcolor[HTML]{FCFCFF}0.31 \\
 FD .5/ H & \cellcolor[HTML]{FCFCFF}0.00    & \cellcolor[HTML]{FCFCFF}0.00    & \cellcolor[HTML]{FCFCFF}0.02 & \cellcolor[HTML]{FCFCFF}0.00    & \cellcolor[HTML]{FCFBFE}0.81 & \cellcolor[HTML]{FCFAFD}1.49 \\
 RS +1k & \cellcolor[HTML]{FCFCFF}0.02 & \cellcolor[HTML]{FCFCFF}0.10  & \cellcolor[HTML]{FCFBFE}0.84 & \cellcolor[HTML]{FCFBFE}1.11 & \cellcolor[HTML]{FCFCFF}0.31 & \cellcolor[HTML]{FCFAFD}1.63
 \end{tabular}%
}%
\end{table}

\begin{table*}[!t]
  \fontsize{6.5pt}{7.8pt}\selectfont
  \renewcommand{\arraystretch}{1.2} 
  \caption{The FARs (\%) of model with different encoder architectures under all manipulations}
  \label{tab:encoder_manipulation_full_appendix}
  \begin{threeparttable}
    \begin{tabular}{l|l|ll|lll|lllllll|llll}
      \hline
    Encoder   & None                        & \multicolumn{2}{c|}{Volume Control}     & \multicolumn{3}{c|}{White Noise} & \multicolumn{7}{c|}{Environmental Noise} & \multicolumn{4}{c}{Time Stretch}                                                                                                  \\
    Architecture       & ~                                        & 0.5                             & 0.1                                   & 15dB                                              & 20dB                             & 25dB  & WD    & FS    & BR    & CO    & RA    & CT    & SN     & 1.1   & 1.05  & 0.95  & 0.9   \\
      \hline 
      RawNet2     & 3.37\phantom{0} & 1.44\phantom{0} & 0.36\phantom{0} & 4.06\phantom{0} & 3.61\phantom{0} & 3.24\phantom{0} & 2.28\phantom{0} & 6.13\phantom{0} & 3.63\phantom{0} & 2.24\phantom{0} & 3.86\phantom{0} & 8.23\phantom{0} & 2.85\phantom{0} & 0.29\phantom{0} & 0.37\phantom{0} & 0.32\phantom{0} & 0.23\phantom{0} \\
      Res-TSSDNet & 3.00 & 1.25 & 0.53 & 0.03 & 0.15 & 0.44 & 0.20 & 0.70 & 3.23 & 0.29 & 0.19 & 0.68 & 1.97 & 0.19 & 0.28 & 0.20 & 0.13 \\
      AASIST      & 1.11 & 0.70 & 0.06 & 0.12 & 0.52 & 0.78 & 1.39 & 1.17 & 0.68 & 0.20 & 0.23 & 1.15 & 1.01 & 0.08 & 0.12 & 0.06 & 0.03 \\
      \hline
    \end{tabular}

    \vspace{1.5em}  

    \begin{tabular}{l|lll|lll|lllllll|llll}
      \hline
      Encoder    & \multicolumn{3}{c|}{Add Echoes} & \multicolumn{3}{c|}{Time Shift} & \multicolumn{7}{c|}{Fade} & \multicolumn{4}{c}{Resample}                                                                                                          \\
      Architecture            & 1k/ .2                                   & 1k/ .5                          & 2k/ .5                                & 1.6k                                              & 16k                              & 32k   & .5/ L & .3/ L & .1/ L & .5/ E & .5/ Q & .5/ H & .5/ Lo & -1k   & -0.5k & +0.5k & +1k   \\
      \hline
      RawNet2     & 2.19\phantom{0} & 1.17\phantom{0} & 1.30\phantom{0} & 2.50\phantom{0} & 3.01\phantom{0} & 3.00\phantom{0} & 1.31\phantom{0} & 2.10\phantom{0} & 2.94\phantom{0} & 2.05\phantom{0} & 1.74\phantom{0} & 7.08\phantom{0} & 1.95\phantom{0} & 3.98\phantom{0} & 3.28\phantom{0} & 2.81\phantom{0} & 2.54\phantom{0} \\
      Res-TSSDNet & 0.71 & 0.14 & 0.16 & 2.41 & 2.43 & 2.48 & 0.70 & 1.08 & 1.61 & 0.52 & 0.99 & 0.86 & 1.30 & 1.78 & 2.19 & 3.43 & 3.87 \\
      AASIST      & 0.21 & 0.03 & 0.07 & 0.85 & 0.93 & 0.89 & 0.64 & 0.95 & 1.06 & 0.50 & 0.85 & 0.81 & 0.93 & 0.65 & 0.81 & 1.38 & 1.63 \\
      \hline
    \end{tabular}
  \end{threeparttable}
  \vspace{-1em}
\end{table*}

\subsection{Manipulation Attacks Results (RQ1)}

To address the first research question, we conducted a large-scale experiment to investigate the impact of manipulations on the performance of the baseline models. Only the audio deepfakes are manipulated to bypass the detection. The results are presented in the first four rows of Tab.~\ref{tab:comparison_under_attacks}.

It is evident that all baseline models exhibit excellent performance in the absence of manipulations, with FARs of 4.60\%, 0.83\%, 1.63\% and 1.09\% for RawNet2, AASIST, Res-TSSDNet and SAMO, respectively. However, simple volume control significantly increases the FAR, reaching 36.69\% for RawNet2 when the control factor is set to 0.1, with similar increases observed for other baselines. 
Regarding noise injection, RawNet2 is vulnerable to white noise, while Res-TSSDNet experiences severe degradation from white noise, wind noise, and rain noise, as indicated by an FAR as high as 51.28\%. Interestingly, AASIST and SAMO performs better, with a lower FAR even compared to the original data, indicating its ability to identify noisy audio as potential deepfakes. 

When facing time stretch manipulation, all models consistently yield better results, likely due to the noticeable artifacts introduced by the FFT and inverse FFT processes in time stretching, which are frequently observed in audio deepfakes. For echoes and timeshift, none of the baselines are affected much. Fading with half sinusoidal fade shape is a strong manipulation that consistently achieving high bypass performance with FARs of 30.43\%, 31.23\%, and 11.55\%, 28.24\% for RawNet2, AASIST, Res-TSSDNet and SAMO, respectively. 
Resampling does not show significant influence, consistent with findings in ~\cite{wang_deepsonar_2020} who tested a narrower range. Regarding F1 scores, we observe similar trends as FARs.

\begin{figure*}[ht]
  \centering
  \includegraphics[width=\textwidth]{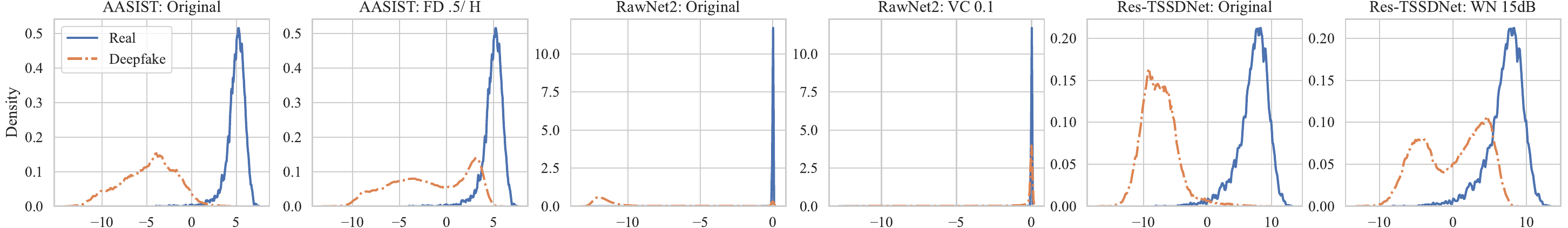}
  \caption{The prediction score distribution of baseline model output for the whole dataset under various types of manipulations.}
  \label{fig:score_dist_manipulations}
\end{figure*}

To examine how manipulation attacks influence the model's prediction score, we take three representative manipualtion attack settings as examples and analyze the score distribution for the entire dataset output by three baselines.
For better visualization, we adopt the approach used by ~\cite{tak_end--end_2021}, which employs the second element of the final linear layer output as the prediction score, which reflects the softmax output.
The distribution of prediction scores for the selected baseline models under various manipulations is presented in Fig.~\ref{fig:score_dist_manipulations}. Notably, higher scores indicate higher level of confidence in the authenticity of the sample. 
We observe that though the original score distribution without manipulation for different models are different, they all share a same trait that the distribution of deepfakes and real audio can be easily separated. However, after manipulation, we observe that the score distribution for the deepfakes shift towards distribution of real audio. Consequently, with the original threshold, we could expect a high FAR for the manipulated samples.

In order to enhance the effectiveness of attacks, a straightforward and intuitive strategy for adversaries is to combine multiple potent manipulations. We assess the performance of baseline models against a combination of manipulation attacks. We employ a representative set of manipulations for evaluation, including: volume control (VC) with a factor of 0.1, white noise (WN) with a SNR of 15 dB, wind environmental noise(EN), time stretch(TS) with a stretching factor of 0.9, fading(FD) with a half sine shape and fade ratio of 0.5, and resampling(RS) to 17,000 Hz.
Tab.~\ref{tab:combination_manipulations} presents the results. Each row represents the first manipulation applied, and each column represents the second. The diagonal elements represent the performance of the baseline models under single manipulation attacks. We observe that RawNet2 and Res-TSSDNet exhibit a significantly higher FAR under combined attacks. In contrast, for AASIST and SAMO, the combination of manipulations does not result in a higher FAR.

To sum up, our evaluation indicates that all baseline models are vulnerable to manipulation attacks. Although the baseline models exhibit varying degrees of vulnerability, volume control, noise injection, and fading generally achieve higher FARs. Furthermore, combining different manipulation techniques results in stronger attack performance for baselines like RawNet2 and Res-TSSDNet.

\subsection{CLAD Performance (RQ2)}

In this section, we evaluate CLAD's effectiveness against manipulation attacks and answer the second research question. Tab.~\ref{tab:comparison_under_attacks} presents the results. We observe a slight increase in FAR from 0.83\% to 1.11\% for CLAD in the absence of manipulation. Notably, 0.83\% represents the best result achieved by AASIST, with the authors reporting an average training result of 1.13\%\cite{jung_aasist_2022}. This increase in FAR is an acceptable trade-off, considering the overall robustness of CLAD. In the presence of white noise injection, the proposed CLAD model outperforms Res-TSSDNet, achieving an FAR of 0.12\%, whereas Res-TSSDNet records a substantially higher FAR of 51.28\%. Similarly, CLAD exhibits significant improvements over RawNet2 and AASIST when subjected to volume control and fading. Even for combinations of manipulations, CLAD maintains a lower FAR compared to using only a single manipulation, as depicted in Tab.~\ref{tab:combination_manipulations}.
In conclusion, CLAD demonstrates robustness against manipulation attacks by maintaining the highest FAR of 1.63\% and the lowest F1 score of 92.82\% among all tested manipulation attack scenarios.

\begin{figure}[!t]
  \centering
  \subfigure[No Manipulation]{\includegraphics[width=0.45\columnwidth]{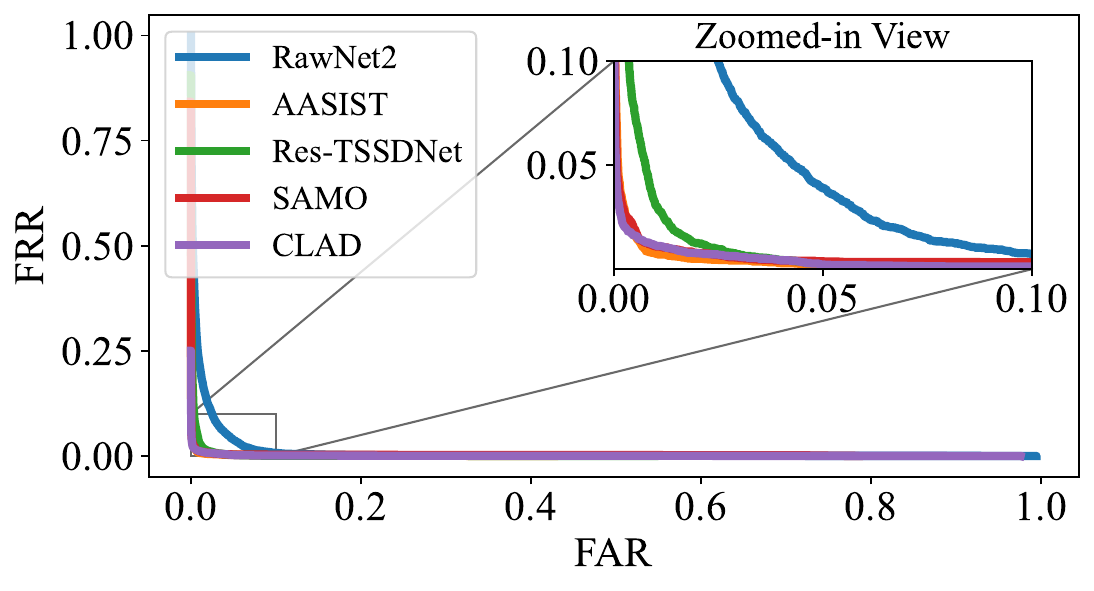}}
  \subfigure[White Noise 15dB]{\includegraphics[width=0.45\columnwidth]{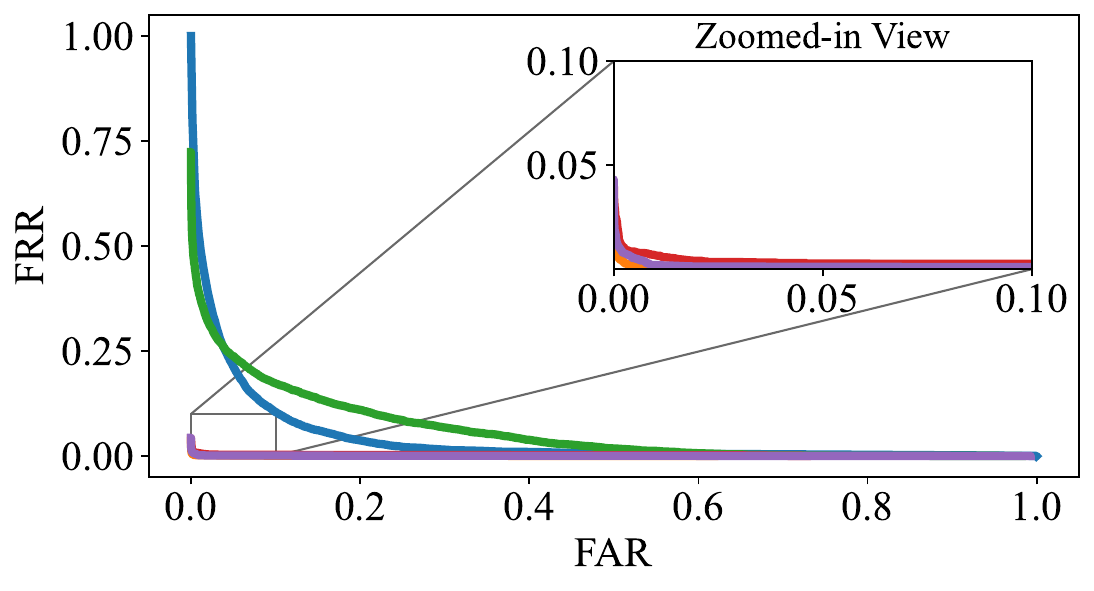}}
  \subfigure[Environmental Noise Rain]{\includegraphics[width=0.45\columnwidth]{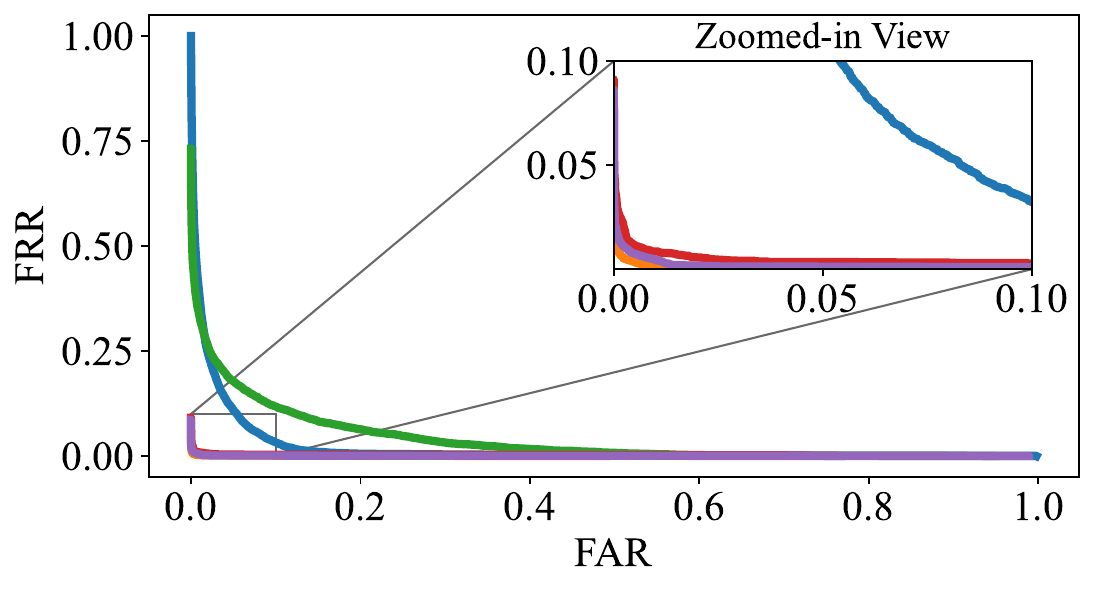}}
  \subfigure[Volume Control 0.1]{\includegraphics[width=0.45\columnwidth]{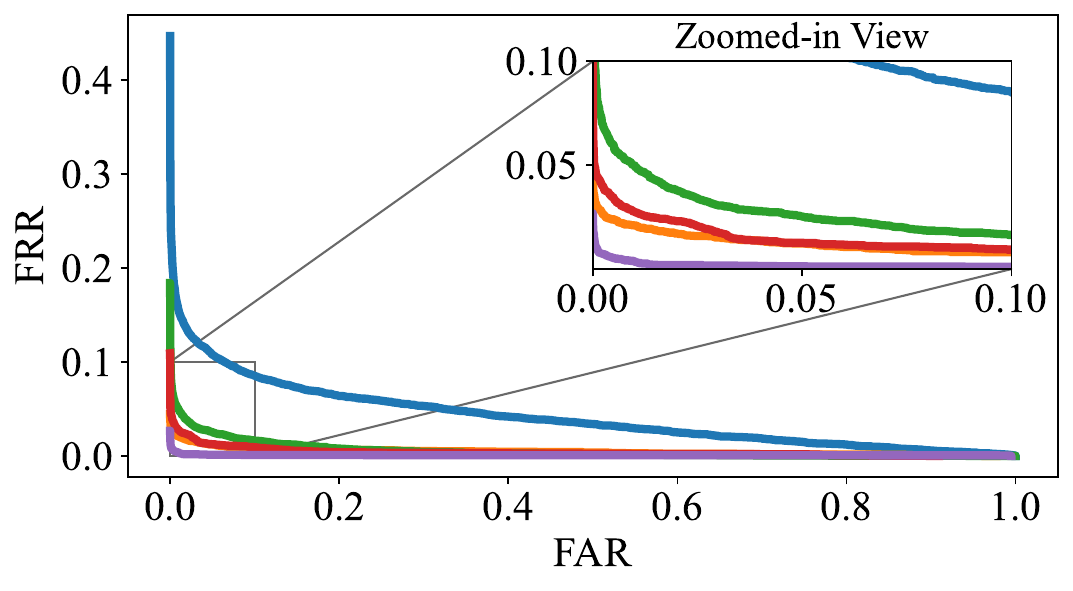}}
  \subfigure[Fade 0.5 Half Sine]{\includegraphics[width=0.45\columnwidth]{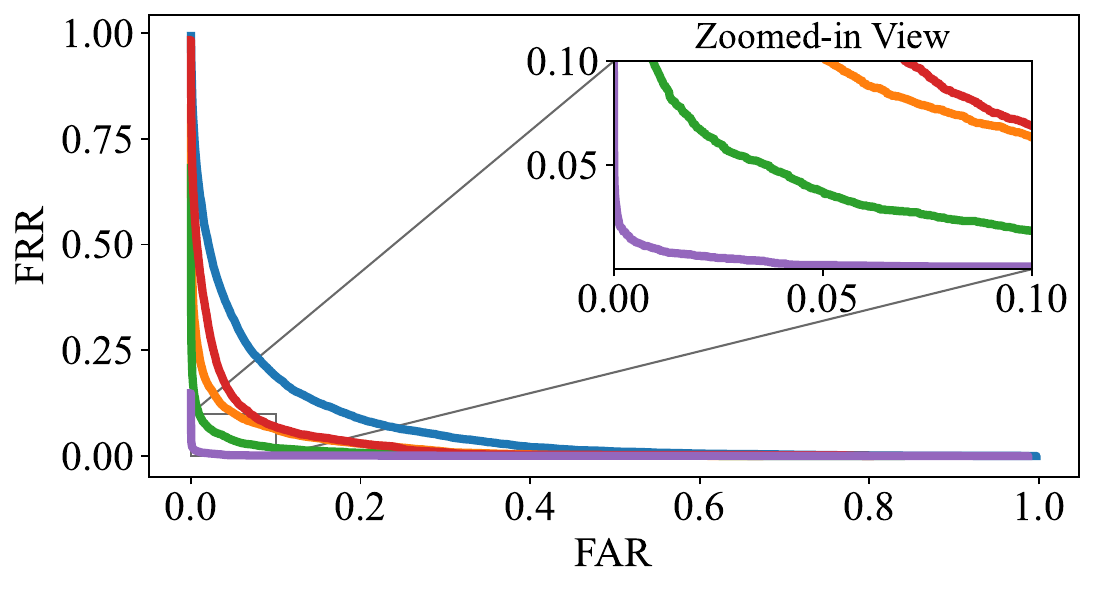}}
  \subfigure[Fade 0.5 Exponential]{\includegraphics[width=0.45\columnwidth]{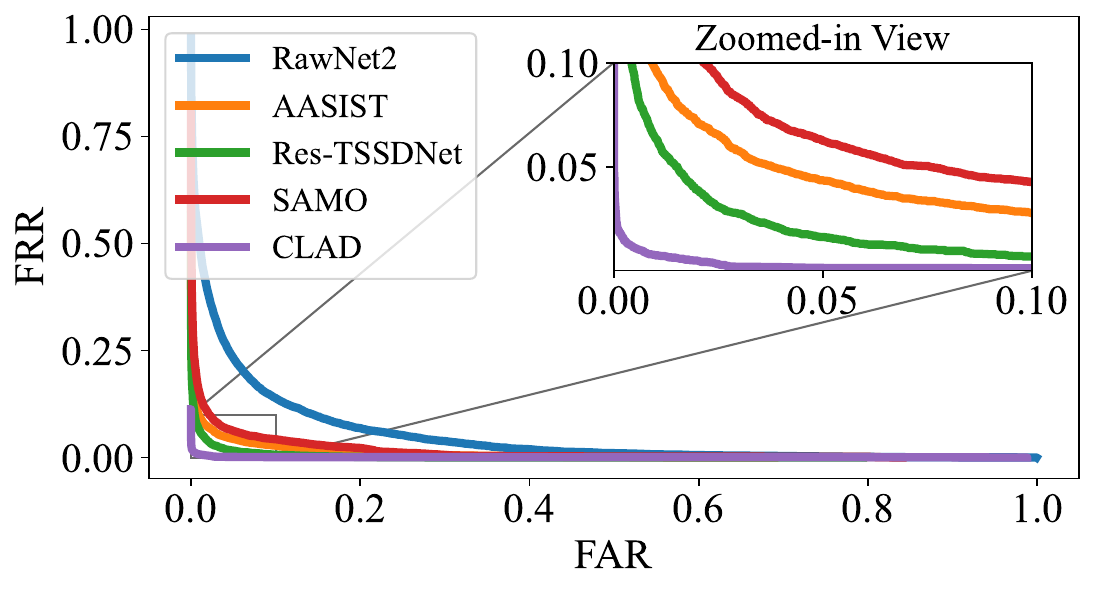}}

  \caption{DET curve of detection models under different manipulations.}
  \label{fig:DET_curve_main_paper}
\end{figure}

Additionally, we present the evaluation results using Detection Error Tradeoff (DET) curves, which offer a comprehensive visualization of the tradeoff between the False Acceptance Rate (FAR) and False Rejection Rate (FRR). Fig.~\ref{fig:DET_curve_main_paper} depicts the DET curves for different models across four manipulation attack settings.
Curves positioned in the lower left quadrant indicate better detection performance. It can be observed that CLAD consistently performs excellently, while the baseline models exhibit varying degrees of performance degradation under certain manipulations.

\begin{figure}[!t]
  \centering
  \includegraphics[width=0.9\columnwidth]{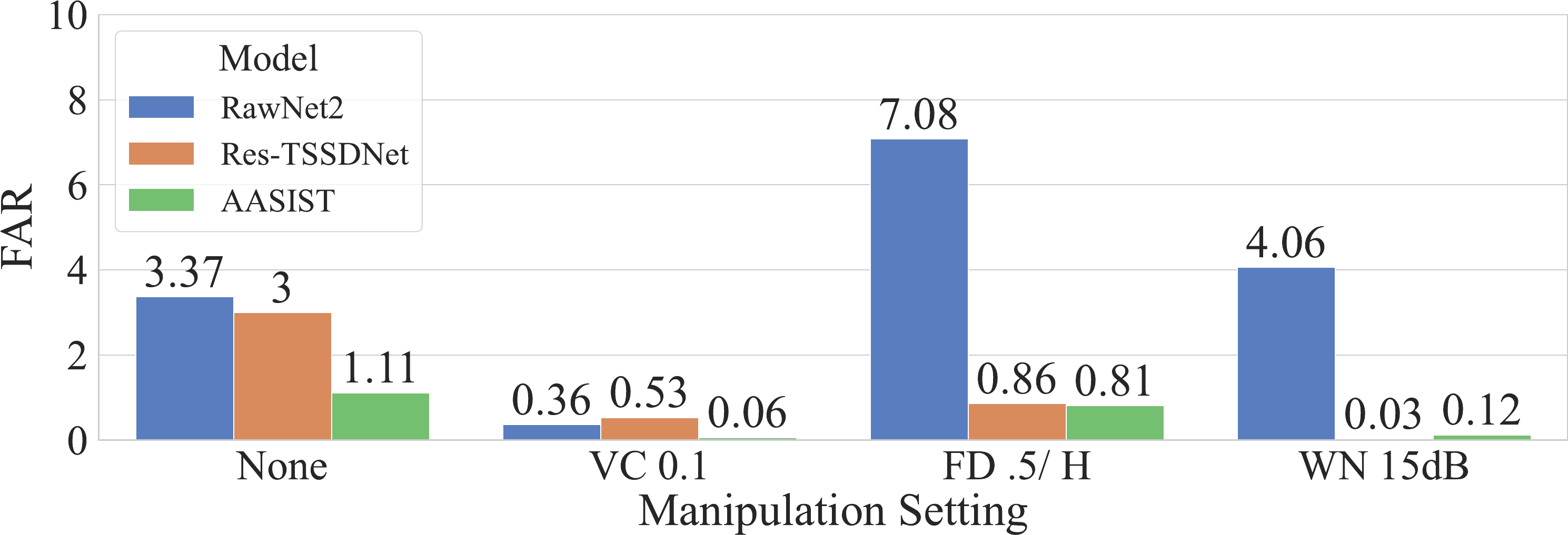}
  \caption{The FARs (\%) of models with differnet encoder architectures under different manipulation attacks}
  \label{fig:FAR_encoder_barplot}
\end{figure}

\begin{figure}[!t]
  \centering
  \includegraphics[width=\columnwidth]{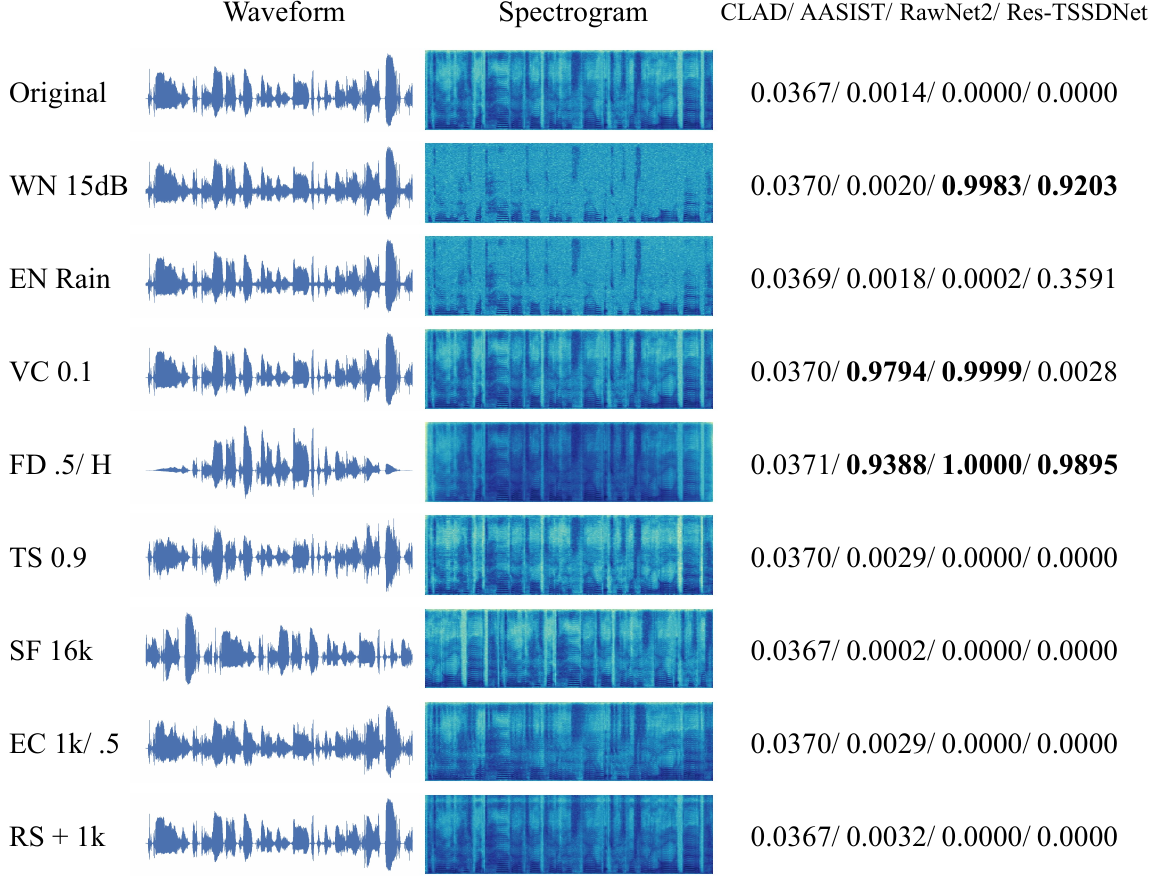}
  \caption{Audio Deepfake examples under various manipulations, along with predicted probabilities of being real. Higher predictions indicate the model mistakenly identifies the audio as real.}
  \label{fig:manipulation_prediction}
\end{figure}

It should be noted that CLAD can be integrated with existing end-to-end detection models by taking them as an encoder. In this study, we selected the AASIST architecture as the encoder for CLAD. To assess the impact of the encoder architecture, we conducted performance evaluations of CLAD using RawNet2 and Res-TSSDNet as alternative encoders. SAMO is based on the same architecture as AASIST, so we do not compare it here. Representative results are illustrated in Fig.~\ref{fig:FAR_encoder_barplot},
and full results can be found in Tab.~\ref{tab:encoder_manipulation_full_appendix}.
The results demonstrate that the encoder architecture do influence the performance of CLAD. Specifically, the model employing an AASIST encoder consistently outperforms the alternative encoders across most experimental conditions. This observation aligns with the baseline model performance reported in Tab.\ref{tab:comparison_under_attacks}, suggesting that a better encoder architecture could enhance the performance of CLAD. 
Moreover, observing the results presented in Tab.~\ref{tab:encoder_manipulation_full_appendix}, it is noteworthy that the performance of CLAD with RawNet2 encoder still outperforms the performance of the original model under most manipulation attacks. We even trained a CLAD model with RawNet2 encoder with better performance on original data compared to the model released by the authors.
Results here underscore the CLAD's effectiveness in improving the robustness of existing deepfake detection models.
Fig.~\ref{fig:manipulation_prediction} presents an illustrative audio deepfake case alongside predictions generated by various detection methods under different manipulation settings for better comprehension.

\begin{table*}[!t]
  \renewcommand{\arraystretch}{1.2} 
  \fontsize{7pt}{8.4pt}\selectfont
  \caption{The FARs (\%) of model variants in ablation study under all manipulations.}
  \label{tab:ablation_full_appendix}
  \begin{threeparttable}
    \begin{tabular}{l|l|ll|lll|lllllll|llll}
      \hline

    Models  & None           & \multicolumn{2}{c|}{Volume Control}     & \multicolumn{3}{c|}{White Noise} & \multicolumn{7}{c|}{Environmental Noise} & \multicolumn{4}{c}{Time Stretch}                                                                                                  \\
      ~       & ~                                        & 0.5                             & 0.1                                   & 15dB                                              & 20dB                             & 25dB  & WD    & FS    & BR    & CO    & RA    & CT    & SN     & 1.1   & 1.05  & 0.95  & 0.9   \\
      \hline
      Vanilla          & 3.02\phantom{0} & 4.80\phantom{00} & 0.19\phantom{0} & 0.12\phantom{0} & 0.71\phantom{0} & 1.67\phantom{0} & 3.19\phantom{0} & 4.12\phantom{0} & 2.87\phantom{0} & 0.99\phantom{0}  & 0.69\phantom{0} & 4.96\phantom{0} & 2.98\phantom{0} & 0.39\phantom{0} & 0.53\phantom{0} & 0.31\phantom{0} & 0.06\phantom{0} \\ 
      CL                & 4.36 & 1.09  & 0.03 & 2.52 & 4.20 & 4.61 & 4.20 & 4.23 & 4.10 & 1.50  & 3.01 & 5.89 & 3.43 & 0.79 & 0.90 & 0.88 & 0.67 \\
      LL             & 2.00 & 13.87 & 2.24 & 6.63 & 4.17 & 2.72 & 4.26 & 5.84 & 4.17 & 13.70 & 4.76 & 7.29 & 4.94 & 9.64 & 7.04 & 7.34 & 9.72 \\
      CLAD                 & 1.11 & 0.70  & 0.06 & 0.12 & 0.52 & 0.78 & 1.39 & 1.17 & 0.68 & 0.20  & 0.23 & 1.15 & 1.01 & 0.08 & 0.12 & 0.06 & 0.03 \\
      \hline
    \end{tabular}

    \vspace{1.5em} 

    \begin{tabular}{l|lll|lll|lllllll|llll}
      \hline
      Models   & \multicolumn{3}{c|}{Add Echoes} & \multicolumn{3}{c|}{Time Shift} & \multicolumn{7}{c|}{Fade} & \multicolumn{4}{c}{Resample}                                                                                                          \\
      ~            & 1k/ .2                                   & 1k/ .5                          & 2k/ .5                                & 1.6k                                              & 16k                              & 32k   & .5/ L & .3/ L & .1/ L & .5/ E & .5/ Q & .5/ H & .5/ Lo & -1k   & -0.5k & +0.5k & +1k   \\
      \hline
      Vanilla & 0.55\phantom{0} & 0.04\phantom{0} & 0.16\phantom{0} & 1.90\phantom{0} & 2.06\phantom{0} & 1.93\phantom{0} & 6.14\phantom{0} & 5.03\phantom{0} & 3.47\phantom{0} & 6.87\phantom{0} & 5.46\phantom{0} & 11.47 & 4.79\phantom{0} & 2.32\phantom{0} & 2.63\phantom{0} & 3.20\phantom{0} & 4.11\phantom{0} \\
      CL       & 1.85 & 0.76 & 1.02 & 3.73 & 3.79 & 3.87 & 1.82 & 3.33 & 4.08 & 1.25 & 2.79 & 1.95  & 3.19 & 3.44 & 3.97 & 4.94 & 5.88 \\
      LL    & 3.73 & 8.12 & 9.46 & 2.79 & 2.92 & 2.90 & 2.52 & 2.15 & 2.04 & 3.33 & 2.26 & 4.43  & 2.09 & 2.34 & 2.26 & 2.22 & 2.54 \\
      CLAD        & 0.21 & 0.03 & 0.07 & 0.85 & 0.93 & 0.89 & 0.64 & 0.95 & 1.06 & 0.50 & 0.85 & 0.81  & 0.93 & 0.65 & 0.81 & 1.38 & 1.63 \\
      \hline
    \end{tabular}
  \end{threeparttable}
\end{table*}

\subsection{Ablation Study (RQ3)}

In order to systematically assess the key components of our approach, we conducted ablation experiments from two perspectives and provide a comprehensive analysis in the following sections.

\begin{table}[!t]\small
  \renewcommand{\arraystretch}{1.2} 
  \caption{The model variants in ablation study}
  \label{tab:ablation_model_vairants_main_paper}
  \centering
  \begin{tabular}{l|cccc}
    \hline
    Component            & \multicolumn{1}{l}{Vanilla} & \multicolumn{1}{l}{CL} & \multicolumn{1}{l}{LL} & \multicolumn{1}{l}{CLAD} \\
    \hline
    Contrastive Learning & \xmark & \cmark & \xmark & \cmark \\
    Length Loss & \xmark & \xmark & \cmark & \cmark \\
    \hline
  \end{tabular}
\end{table}

\textbf{Model Variants}.
We consider four model variants to evaluate the components of CLAD, which are summarized in Tab.~\ref{tab:ablation_model_vairants_main_paper}.
The ablation study results are presented in Tab.~\ref{tab:ablation_full_appendix}.

\begin{enumerate}[leftmargin=*]
  \item Vanilla. This is an AASIST model trained without contrastive learning and length loss. Compared with the baseline, it employs all manipulations as data augmentation.
  \item CL. This represents our model trained without length loss.
  \item LL. Refers to the Vanilla model trained with the inclusion of length loss as an additional loss function.
  \item CLAD. Denotes our proposed method, which utilizes both contrastive learning and length loss.
\end{enumerate}

\textbf{Effectiveness of Contrastive Learning}.
Examining Tab.~\ref{tab:ablation_full_appendix}, it is evident that contrastive learning proves effective in enhancing the model's robustness against manipulation attacks. For models trained without contrastive learning, i.e., Vanilla and LL, we observe FAR values of 11.47\% and 13.87\% under fading and volume control, respectively. Even though these models are trained with all the manipulations, conventional supervised learning struggles to handle such diverse manipulations effectively. In contrast, CL and CLAD yield more consistent results across all manipulations. It is worth noting that the FAR of LL for the original dataset stands at 4.36\%, which is less favorable compared to other models. This discrepancy is attributed to the unsupervised nature of contrastive learning, which results in the model being insufficiently trained with label information.  

\begin{figure}[!t]
  \centering
  \subfigure[w/o length loss]{\includegraphics[width=0.48\columnwidth]{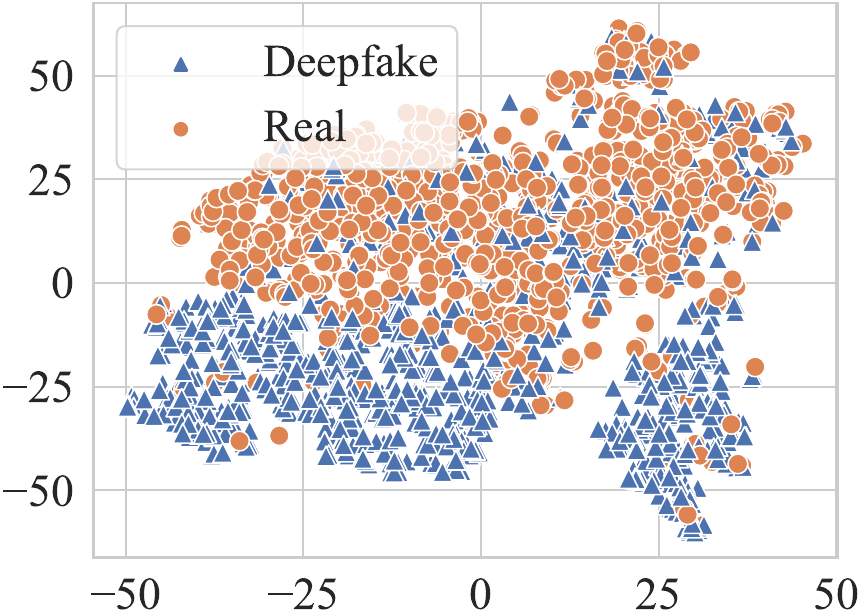}}
  \subfigure[w/ length loss]{\includegraphics[width=0.48\columnwidth]{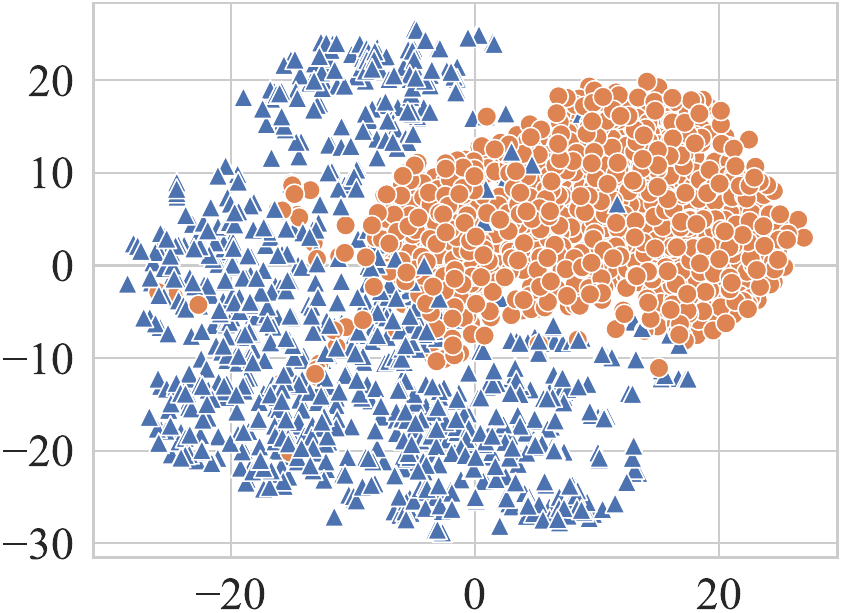}}
  \caption{Visualization of the features extracted using the encoders trained without and with length loss.}
  \label{fig:feature_distribution}
  \vspace{-1em}
\end{figure}

\textbf{Effectiveness of Length Loss}.
Length loss encourages the model to learn the clustering of real audio, thus making downstream training easier. As demonstrated in Tab.~\ref{tab:ablation_full_appendix}, models trained with length loss outperform those without it on the original data. LL achieves an FAR of 2.00\%, while Vanilla achieves 3.02\%. Notably, CLAD shows a substantial improvement compared to CL, with CLAD achieving a FAR of 1.11\% compared to CL's 4.36\%. Similar improvement can also be observed under various manipulations. Such enhancements can be attributed to the design of the length loss, which utilizes length of feature vectors and complements contrastive learning.
To provide further insight into the impact of length loss, we visualize the feature vectors extracted by the encoder, both trained with and without length loss, in Fig.~\ref{fig:feature_distribution}. The visualization illustrates that the encoder trained with length loss produces feature vectors that are more closely aggregated. This, in turn, enhances detection performance.

\begin{table}[t!]
  \caption{The FARs (\%) of models trained with partial manipulations under different manipulation attacks including unknown manipulation.}
  \label{tab:FAR_heatmap_unknown_manipulation}
  \centering
  \scriptsize  
  \renewcommand{\arraystretch}{1.2}
  \centering
  \begin{tabular}{lllllllll}
    \rowcolor[HTML]{ECECEC} 
    & \multicolumn{1}{l}{VC}       & \multicolumn{1}{l}{WN}       & \multicolumn{1}{l}{EN}       & \multicolumn{1}{l}{TS}       & \multicolumn{1}{l}{EC}       & \multicolumn{1}{l}{SF}       & \multicolumn{1}{l}{FD}       & \multicolumn{1}{l}{RS}       \\
None               & \cellcolor[HTML]{FCDFE2}1.49 & \cellcolor[HTML]{FBC9CC}2.61 & \cellcolor[HTML]{FBD6D9}1.96 & \cellcolor[HTML]{FBD5D7}2.03 & \cellcolor[HTML]{FCDADD}1.76 & \cellcolor[HTML]{FCE4E7}1.26 & \cellcolor[HTML]{FBD0D3}2.24 & \cellcolor[HTML]{FBD3D6}2.11 \\
VC 0.1 & \cellcolor[HTML]{F8696B}7.39 & \cellcolor[HTML]{FBC6C8}2.77 & \cellcolor[HTML]{FBD4D7}2.05 & \cellcolor[HTML]{FCDBDE}1.70 & \cellcolor[HTML]{FBC3C6}2.90 & \cellcolor[HTML]{FCDBDE}1.69 & \cellcolor[HTML]{FCF3F6}0.49 & \cellcolor[HTML]{FAA8AB}4.25 \\
WN 15dB   & \cellcolor[HTML]{FCF3F6}0.50 & \cellcolor[HTML]{FCE7E9}1.12 & \cellcolor[HTML]{FCFAFD}0.16 & \cellcolor[HTML]{FCF3F6}0.52 & \cellcolor[HTML]{FCF5F8}0.40 & \cellcolor[HTML]{FCF9FC}0.19 & \cellcolor[HTML]{FCF6F9}0.37 & \cellcolor[HTML]{FCF6F9}0.36 \\
EN Rain     & \cellcolor[HTML]{FCEDF0}0.80 & \cellcolor[HTML]{FCDCDF}1.65 & \cellcolor[HTML]{FCF1F4}0.59 & \cellcolor[HTML]{FCEAED}0.93 & \cellcolor[HTML]{FCECEF}0.83 & \cellcolor[HTML]{FCF1F4}0.58 & \cellcolor[HTML]{FCECEF}0.84 & \cellcolor[HTML]{FCEBEE}0.90 \\
TS 0.9  & \cellcolor[HTML]{FCF8FB}0.25 & \cellcolor[HTML]{FCF3F6}0.52 & \cellcolor[HTML]{FCF7FA}0.28 & \cellcolor[HTML]{FCFCFF}0.03 & \cellcolor[HTML]{FCF6F9}0.34 & \cellcolor[HTML]{FCFCFF}0.03 & \cellcolor[HTML]{FCF9FC}0.21 & \cellcolor[HTML]{FCF9FC}0.22 \\
EC 1k/ .2          & \cellcolor[HTML]{FCF0F3}0.67 & \cellcolor[HTML]{FCE5E8}1.19 & \cellcolor[HTML]{FCEEF1}0.73 & \cellcolor[HTML]{FCF0F3}0.67 & \cellcolor[HTML]{FCF7FA}0.28 & \cellcolor[HTML]{FCF7FA}0.30 & \cellcolor[HTML]{FCEAED}0.96 & \cellcolor[HTML]{FCF3F6}0.50 \\
SF 16k     & \cellcolor[HTML]{FCE5E8}1.21 & \cellcolor[HTML]{FBD1D4}2.20 & \cellcolor[HTML]{FCDEE1}1.55 & \cellcolor[HTML]{FCDADD}1.74 & \cellcolor[HTML]{FCE0E3}1.45 & \cellcolor[HTML]{FCECEF}0.87 & \cellcolor[HTML]{FBD6D8}1.97 & \cellcolor[HTML]{FCD9DC}1.81 \\
FD .5/ H          & \cellcolor[HTML]{FBC5C8}2.81 & \cellcolor[HTML]{FBBDBF}3.22 & \cellcolor[HTML]{FBCFD2}2.30 & \cellcolor[HTML]{FBBBBD}3.32 & \cellcolor[HTML]{FBB5B8}3.60 & \cellcolor[HTML]{FBCED1}2.34 & \cellcolor[HTML]{F98184}6.19 & \cellcolor[HTML]{FBD5D8}2.00 \\
RS +1k       & \cellcolor[HTML]{FCDADC}1.77 & \cellcolor[HTML]{FBB7B9}3.52 & \cellcolor[HTML]{FBC6C9}2.75 & \cellcolor[HTML]{FBCFD2}2.31 & \cellcolor[HTML]{FBCFD1}2.32 & \cellcolor[HTML]{FCDBDD}1.72 & \cellcolor[HTML]{FBC3C6}2.90 & \cellcolor[HTML]{FBC0C3}3.04
\end{tabular}%
\vspace{-1em}
\end{table}

\subsection{Unknown Manipulation Study (RQ4)}

In this section, we assess the detection capability of CLAD for unknown manipulation attack methods, considering that the manipulation methods discussed in our paper may not encompass all methods used by attackers. Of the eight manipulation methods examined in our study, we trained eight CLAD models, each with one manipulation method removed. Subsequently, we assessed the performance of these models when confronted with unknown manipulation attacks. The representative results are presented in Tab.~\ref{tab:FAR_heatmap_unknown_manipulation}, each column represents a model trained without a specific manipulation.

We observed that volume control and fading are particularly strong attacks, as models trained without these manipulations performed not well, resulting in a FAR of 7.19\% and 6.19\%, respectively. However, for other manipulations, models trained without them still achieved good performance under unknown manipulation attacks. For instance, the model trained without time stretch achieved a low FAR of 0.03\% under time stretch manipulation. Similar results were observed for time shift and echoes adding. 
Moreover, compared to baseline models, CLAD demonstrated improved performance against unknown manipulations in most scenarios. For example, although both trained without fading manipulation, our design enabled CLAD to reduce its FAR from 31.22\% to 6.19\% compared with baseline AASIST model. 
Therefore, we conclude that while strong manipulations like volume control and fading may lead to suboptimal results, CLAD is expected to outperform baseline models and achieve good performance against most unknown manipulation attacks.

\subsection{Key Findings}  

\begin{itemize}[leftmargin=*]
  \item \textbf{Manipulation attacks pose a severe threat to deepfake detection models.} We found that all the baseline models we evaluated are vulnerable to particular manipulations. Among manipulations, volume control and fading demonstrate the most favorable attack performance. While volume control can be mitigated through straightforward techniques like input normalization, fading manipulations pose a significant challenge to current detection methods. Fading with a half sine fade shape demonstrates a notably high 25.36\% average FAR across the four baseline models.
  \item \textbf{CLAD enhances model resilience against manipulations.}  We successfully trained a robust audio deepfake detection model with AASIST encoder architecture, achieving a low 1.63\% FAR under all manipulations. Furthermore, our experiments confirm that CLAD can serve as a plug-and-play module to enhance the robustness of existing deepfake detection models. Notably, CLAD trained with the RawNet2 encoder outperforms the model released by the authors, even in unmanipulated data.
  \item \textbf{Contrastive Learning and Length Loss are essential components of CLAD.} We found that contrastive learning trained model perform more consisitently under different manipulation attacks, indicating its improved robustness. Length loss is also an essential component of CLAD, complementing contrastive learning and further improving its performance. Moreover, our evaluation results suggest that intuitive data augmentation is not the optimal solution for countering manipulation attacks.
  \item \textbf{CLAD shows promise for superior performance against unknown manipulation attacks.} Our study reveals that CLAD demonstrates promising performance in the presence of unknown manipulation attacks. Even against strong manipulations like volume control and fading, it outperforms baseline models.
\end{itemize}

\section{Related Work}

\subsection{Contrastive Learning}

Contrastive learning is a popular form of self-supervised learning~\cite{liu_self_supervised_2023, jing_self_supervised_2021} that encourages augmentations (views) of the same input to have more similar representations than augmentations of different inputs. Earlier studies ~\cite{chen_simple_2020,he_momentum_2020} illustrates the effectiveness of contrast learning by showing that contrastive learning can even yield better results than supervised learning in some cases. Although contrastive learning has been a great success in the field of computer vision~\cite{ kang_contragan_2020, park_fair_2022, wang_contrastive_2021}, it has not received enough attention in the audio domain.
Saeed et al. ~\cite{saeed_contrastive_2021} proposed COLA, a self-supervised pre-training approach for learning a general-purpose representation of audio. However, COLA does not discuss the downstream tasks of audio deepfake detection. 
Guan et al. ~\cite{guan_anomalous_2023} introduced contrastive learning into the anomalous sound detection and showed its effectiveness. In conclusion, existing work on contrast learning is not sufficient for the audio domain, and as far as we know, our work is the first to apply contrast learning to the field of audio deepfake detection.

\subsection{Audio Deepfake Detection}

The rise of audio deepfakes has posed significant security threats, prompting a surge of interest in developing robust detection methods~\cite{yi_add_2022, yi_add_2023, frank_wavefake_2021}.
Earlier approaches~\cite{todisco_constant_2017,lavrentyeva_stc_2019, ahmed_void_2020} were mainly based on handcrafted features and machine learning-based classifiers. In 2021, Tak et al.~\cite{tak_end--end_2021} proposed a milestone model which is based on an end to end speaker verification model, RawNet2~\cite{jung_improved_2020}. Remarkably, RawNet2 demonstrated state-of-the-art (SOTA) performance without the need for hand-crafted features. Its pre-trained models quickly gained popularity within the field. After that, Jung et al. ~\cite{jung_aasist_2022} proposed a heterogeneous stacking graph attention layer that models artefacts spanning heterogeneous temporal and spectral intervals, and achieved SOTA performance on the ASVspoof 2019 dataset. Blue et al. ~\cite{blue_who_2022} leverage fluid dynamics to estimate the human vocal tract arrangement during speech production and use it to identify deepfakes. This approach, however, needs accurate phoneme timestamps, which are seldom accessible in practical scenarios.

Recently, instead of solely focusing on benchmark datasets performance, researchers have begun considering broader aspects of audio deepfake detection. Recognizing the limitations of existing models on non-English data, Ba et al.~\cite{ba_transferring_2023} proposed domain adaptation strategies for cross-lingual detection. Zhang et al.~\cite{zhang_you_2023, zhang_what_2024} investigated continual learning in audio deepfake detection, which is an effective way to help detection models adapt to new attack types. Zhang et al.~\cite{zhang_compressed_2023} and Singh et al.~\cite{singh_yadav_assd_2023} investigated the challenges in detecting audio deepfakes in compressed form.

However, most of these approaches do not consider the scenarios where the attacker manipulate the speech. A notable exception is DeepSonar ~\cite{wang_deepsonar_2020}, which evaluates the robustness of the model and performs a small-scale experiment of manipulation attacks. In this work, we conduct a larger scale experiment on ASVSpoof 2019 evaluation set (containing 71,237 samples) with more types of manipulation attacks.

\section{Discussion}

\textbf{Intuitive Defense}. An intuitive defense against manipualtion attacks is denoising, as noise injection is one of manipulations. However, our preliminary experiment with Wiener filtering does not yield satisfactory outcomes. While modern speech enhancement methods have potentials, we are concerned that enhancements artefact might trigger false alarms in the detection model. We leave further investigation into this matter to future research.

\textbf{Diverse Datasets}. In this study, we opt to evaluate the baselines using the parameters provided by the authors to make the results more convincing. However, this choice limited our dataset to the ASVspoof 2019 dataset, as the pretrained models did not perform well on other publicly available datasets. With the advancement of the field, we anticipate that future models will exhibit improved generalization and adaptable to a wider range of datasets, enabling evaluations on a more diverse dataset in the future.

\section{Conclusion}

In this paper, we expose the threat of manipulation attacks against audio deepfake detection systems.
It does not require prior knowledge of the system, making it simple for deepfake creators to degrade the detection system performance.
To address this issue, we conduct a large-scale evaluation of widely adopted models on the ASVspoof dataset under manipulation attacks
and find that existing SOTA models experience severe performance degradation when faced with manipulation attacks.
To mitigate the impact of manipulation attacks, we propose contrastive learning-based detection approach and design length loss to train a model to produce robust representations against manipulations.
Experimental results show that CLAD achieves significantly better performance than the widely adopted models under manipulation attacks.

\bibliographystyle{IEEEtran}
\bibliography{library}

\vspace{-33pt}
\begin{IEEEbiography}[{\includegraphics[width=1in,height=1.25in,clip,keepaspectratio]{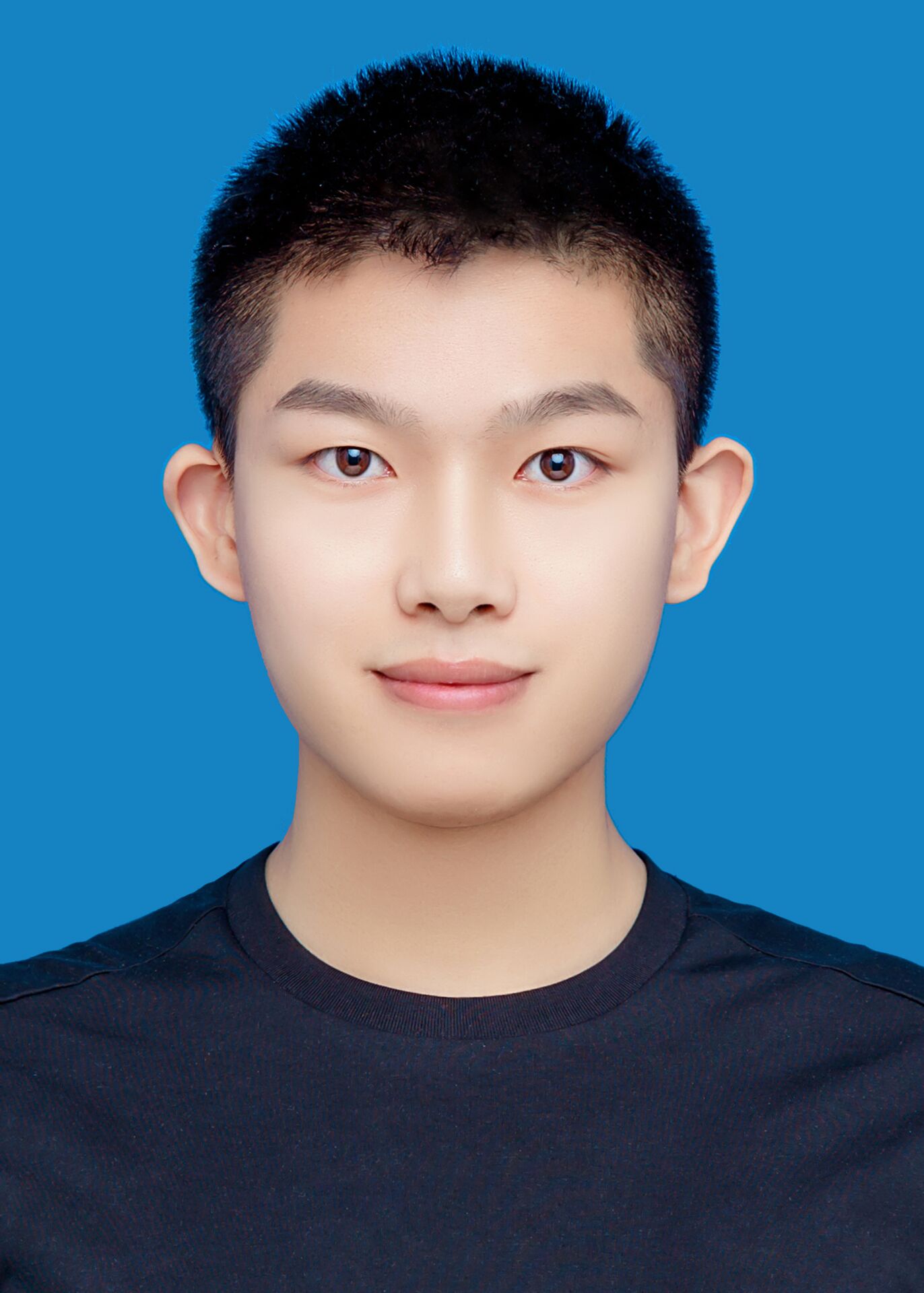}}]{Haolin Wu}
  received the B.E. degree in Information Security from from Wuhan University, Hebei, China, in 2021.
	He is currently pursuing the Ph.D. degree with the School of Cyber Science and Engineering, Wuhan University,
	Hubei, China. His research interest lies in the area of multimedia and machine learning security and privacy.
\end{IEEEbiography}
\vspace{-33pt}
\begin{IEEEbiography}[{\includegraphics[width=1in,height=1.25in,clip,keepaspectratio]{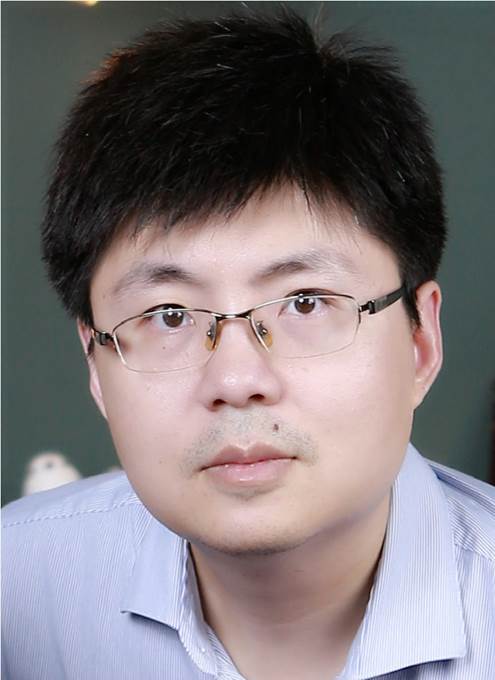}}]{Jing Chen}
  received the Ph.D. degree in computer science from Huazhong University of Science and Technology, Wuhan, China.
  He is the deputy Dean of the School of Cyber Science and Engineering at Wuhan University.
	His research interests are in the areas of network security, cloud security, and mobile security.
	He has published more than 100 research papers in many international journals and conferences, including USENIX Security, ACM CCS, INFOCOM, IEEE TDSC, IEEE TIFS, IEEE TMC, IEEE TC, IEEE TPDS, IEEE TSC, etc.
	He acts as a reviewer for many conferences and journals, such as IEEE INFOCOM, IEEE Transactions on Information Forensics and Security, IEEE Transactions on Dependable and Secure Computing and IEEE/ACM Transactions on Networking.
\end{IEEEbiography}
\vspace{-33pt}
\begin{IEEEbiography}[{\includegraphics[width=1in,height=1.25in,clip,keepaspectratio]{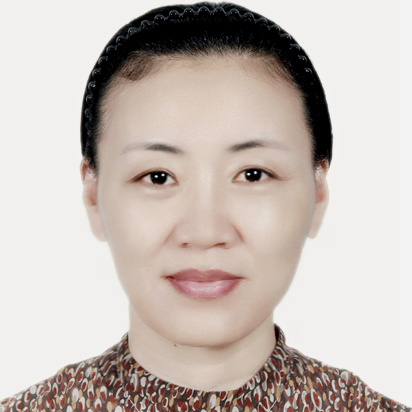}}]{Ruiying Du}
  received the B.S., M.S., and Ph.D. degrees in computer science in 1987, 1994 and 2008, from Wuhan University, Wuhan, China. She is currently a Professor with the School of Cyber Science and Engineering, Wuhan University. Her research interests include network security, wireless network, cloud computing and mobile computing. She has published more than 80 research papers in many international journals and conferences, such as TPDS, USENIX Security, CCS, INFOCOM, SECON, TrustCom, NSS.
\end{IEEEbiography}
\vspace{-33pt}
\begin{IEEEbiography}[{\includegraphics[width=1in,height=1.25in,clip,keepaspectratio]{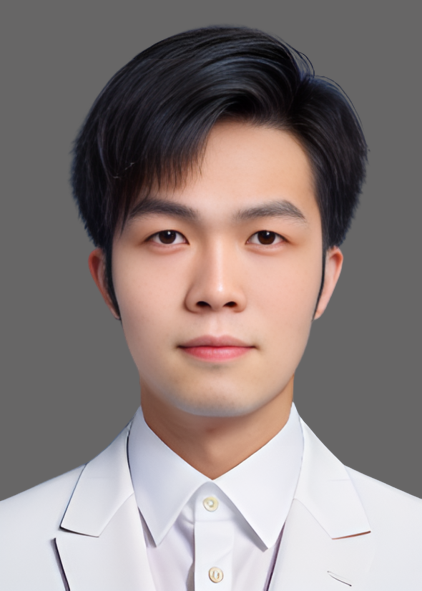}}]{Cong Wu}
  is currently a Research Fellow at Cyber Security Lab, Nanyang Technological University, Singapore. He received Ph.D. degree at School of Cyber Science and Engineering, Wuhan University in 2022. His research interests include AI system security and Web3 security. His leading research outcomes have appeared in USENIX Security, ACM CCS, IEEE TDSC, TMC.
\end{IEEEbiography}
\vspace{-33pt}
\begin{IEEEbiography}[{\includegraphics[width=1in,height=1.25in,clip,keepaspectratio]{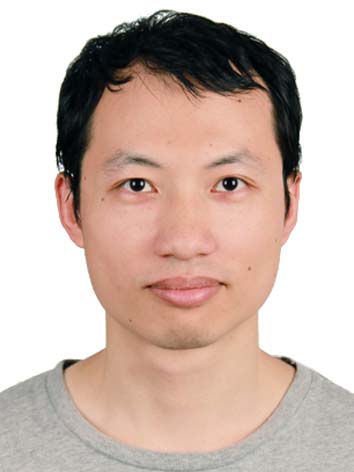}}]{Kun He}
  received his Ph.D. from Wuhan University, Wuhan, China. He is currently an associate professor with Wuhan University. His research interests include cryptography and data security. He has published more than 30 research papers in various journals and conferences, such as TIFS, TDSC, TMC, USENIX Security, CCS, and INFOCOM.
\end{IEEEbiography}
\vspace{-33pt}
\begin{IEEEbiography}[{\includegraphics[width=1in,height=1.25in,clip,keepaspectratio]{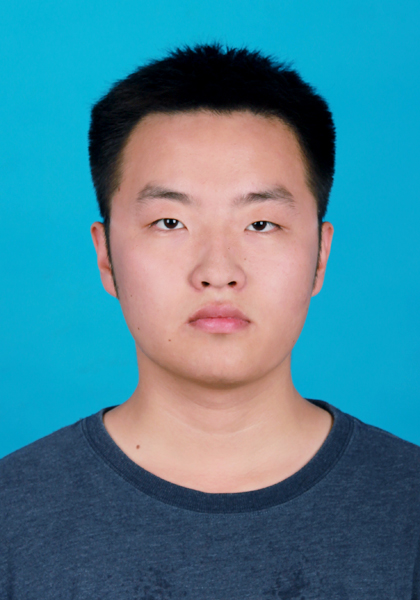}}]{Xingcan Shang}
  received the B.S. degree in computer science and technology from Central South University, Hunan, China, in 2017. He is currently pursuing the Ph.D. degree with the School of Cyber Science and Engineering, Wuhan University, Wuhan, China. His research interest include artificial intelligence security.
\end{IEEEbiography}
\vspace{-33pt}
\begin{IEEEbiography}[{\includegraphics[width=1in,height=1.25in,clip,keepaspectratio]{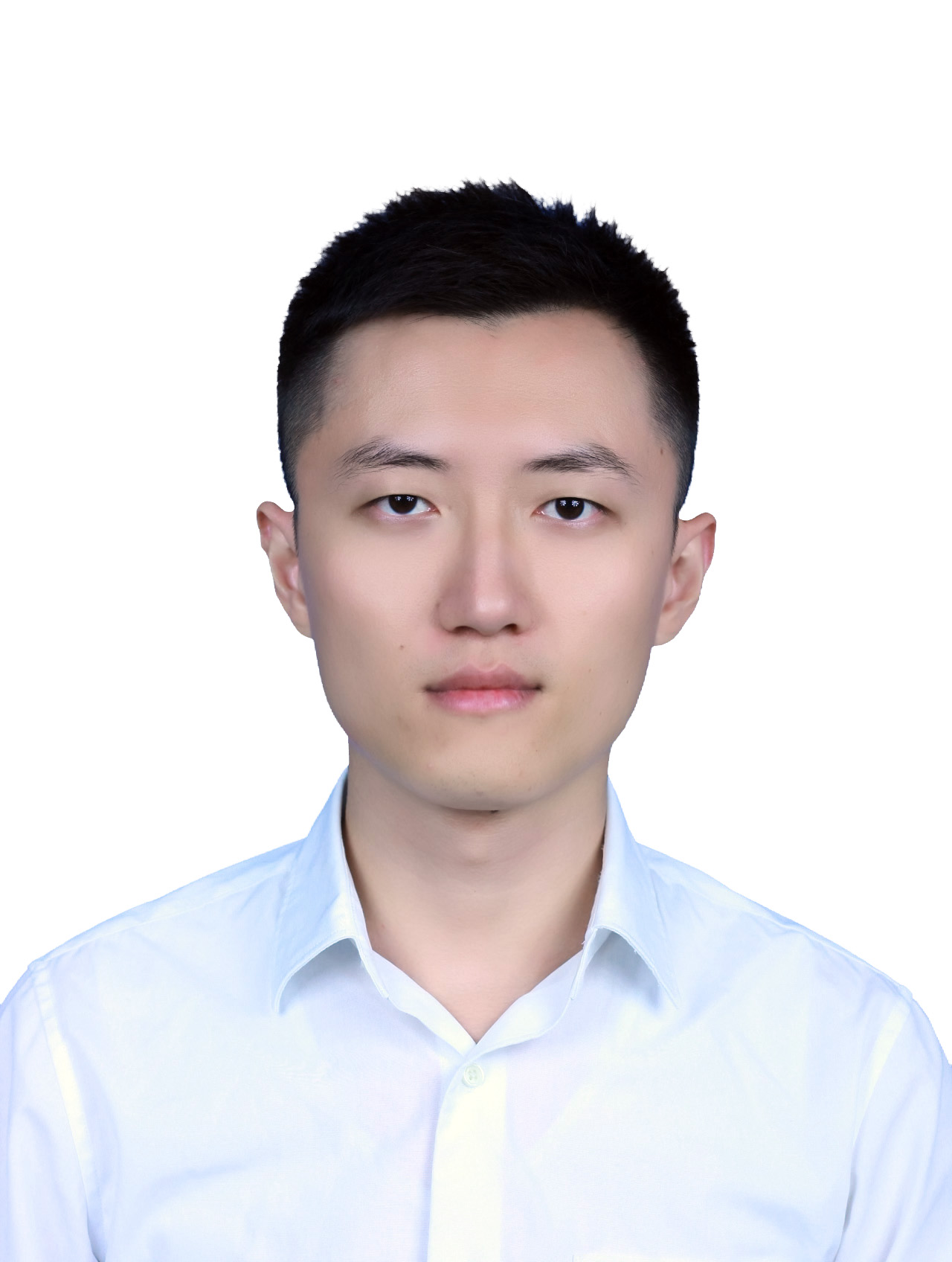}}]{Hao Ren}
  is currently a research associate professor at the Sichuan University. He was a research fellow at Nanyang Technological University from Jul. 2022 to Feb. 2024, at The Hong Kong Polytechnic University from Aug. 2021 to Jun. 2022. He received his Ph.D. degree in Dec. 2020 from the University of Electronic Science and Technology of China. He was a visiting Ph.D. student at the University of Waterloo from Dec. 2018 to Jan. 2020. His research outcomes appeared in major conferences and journals, including WWW, ACM ASIACCS, ACSAC, IEEE TCC, and IEEE Network. He won the Best Paper Award from IEEE BigDataSecurity 2023. His research interests include data security and privacy, applied cryptography and privacy-preserving machine learning.
\end{IEEEbiography}
\vspace{-33pt}
\begin{IEEEbiography}[{\includegraphics[width=1in,height=1.25in,clip,keepaspectratio]{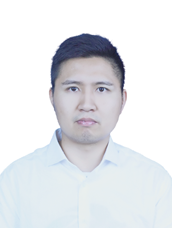}}]{Guowen Xu}
  is currently a Postdoc at City University of Hong Kong. He obtained his Ph.D. degree from University of Electronic Science and Technology of China in 2020. His research focuses on applied cryptography and privacy-preserving deep learning, resulting in over 80 publications in reputable security conferences/journals including IEEE S\&P, ACM CCS, ICML, NeurIPS, ICLR, CVPR, TDSC and TIFS. He was recognized as one of Stanford World's Top 2\% Scientists in 2023.
  Currently, He serves as the Associate Editor for IEEE TIFS, IEEE TNSM, Lead Guest Editor of ACM TAAS, and hold the title of Distinguished Reviewer for ACM TWEB. Moreover, he has had the privilege of participating as a TPC member for esteemed conferences such as ICML (Area Chiar 2024), ACSAC 2024, NeurIPS 2023, CVPR 2023, WWW 2022, AAAI 2022-2023, KSEM 2022, and ICC 2024.
\end{IEEEbiography}
  
\appendix

\section*{Manipulation Details}
\label{sec:appendix_manipulation_details}

In the following section, we provide a description of the manipulation process and define the parameters used. For more detailed information, readers are encouraged to refer to our implementation(\url{https://github.com/CLAD23/CLAD}).

In the case of noise injection, we generate noise and adjust its amplitude to achieve a desired signal-to-noise ratio (SNR). A higher SNR corresponds to a lower level of added noise. In the context of environmental noise injection, we empirically identified the most effective environmental noise source from the ESC-50 dataset.

Volume control involves multiplying the amplitude of the audio waveform by a specified factor. The factor dictates the extent and direction of volume adjustment, whether it entails an increase or decrease, and the magnitude of the change.

Fading involves applying a specific fade shape to attenuate portions of audio at both the beginning and end. The fade shape defines a mask which is applied as a multiplier to the audio's amplitude. The fade ratio determines the extent of audio fading. For instance, a ratio of 0.5 implies that half of the audio at the beginning and half at the end will be faded. The maximum selectable ratio is also 0.5. Detailed information regarding the fade shape definition can be found in the PyTorch implementation.

Time stretching is performed using the official PyTorch implementation. This process alters the audio signal's duration while preserving its pitch. The FFT length is utilized in the FFT calculations involved. The time stretching factor specifies the ratio between the length of the stretched audio and the original audio. A factor less than 1 speeds up the audio.

Resampling is conducted using the official PyTorch implementation. This operation adjusts the audio signal's sample rate to a specified target rate. The target resampling rates defines the sample rate after resampling. When resampled audio is played at the original sample rate, both pitch and duration are affected.

Time shifting involves shifting the audio signal in the time domain by moving all samples forward or backward. The shift length specifies both the direction of audio signal movement and the number of samples shifted.

Echoes adding creates a delayed and attenuated copy of the original audio and adds them together. The delay parameter determines the echo's delay in samples, and the attenuation factor represents the number by which the echo's amplitude is multiplied.

All manipulations were implemented to process the raw waveform and output the raw waveform.

\end{document}